%% file: paper.tex
\documentclass[sigconf]{acmart} 

\usepackage{amsmath,amsfonts}
\usepackage{algorithmic}
\usepackage{graphicx}
\usepackage{textcomp}
\usepackage{xcolor}
\usepackage{subfiles}
\usepackage{tabularx}
\usepackage{booktabs}
\usepackage[capitalise]{cleveref}
\usepackage{xspace}
\usepackage{url}
\usepackage[justification=centering]{caption}
\usepackage{subcaption}
\usepackage{balance}
\usepackage{tabularx}
\usepackage{bm}
\usepackage{listings}
\usepackage{float}
\usepackage{cleveref}
\usepackage{csquotes}
\usepackage[colorinlistoftodos,prependcaption,textsize=tiny]{todonotes}
\graphicspath{{./}{./img/}}

\newcommand{\toolname}{\textsc{Gamekins}\xspace}
\newcommand{\university}{University of Passau\xspace}

\makeatletter
\newcommand\footnoteref[1]{\protected@xdef\@thefnmark{\ref{#1}}\@footnotemark}
\makeatother

\newcommand{\summary}[2]{%
	\vspace{-0.2cm}%
	\begin{center}%
		\colorbox{gray!20}{%
			\parbox{\linewidth}{%
				\textbf{\textsf{Summary (\textit{#1})}:}~%
				#2%
			}%
		}%
	\end{center}%
}

\crefname{lstlisting}{listing}{listings}
\Crefname{lstlisting}{Listing}{Listings}

\copyrightyear{2024}
\acmYear{2024}
\setcopyright{acmlicensed}\acmConference[ICSE-SEET '24]{46th International Conference on Software Engineering: : Software Engineering Education and Training}{April 14--20, 2024}{Lisbon, Portugal}
\acmBooktitle{46th International Conference on Software Engineering: : Software Engineering Education and Training (ICSE-SEET '24), April 14--20, 2024, Lisbon, Portugal}
\acmDOI{10.1145/3639474.3640054}
\acmISBN{979-8-4007-0498-7/24/04}

\begin{document}
	
	\title{Gamifying a Software Testing Course \\with Continuous Integration}
	
	\author{Philipp Straubinger}
	\affiliation{%
		\institution{University of Passau}
		\country{Germany}}
	
	\author{Gordon Fraser}
	\affiliation{%
		\institution{University of Passau}
		\country{Germany}}
	
	\renewcommand{\shortauthors}{Straubinger et al.}
	
	\begin{abstract}
		Testing plays a crucial role in software development, and it is essential for software engineering students to receive proper testing education. However, motivating students to write tests and use automated testing during software development can be challenging. To address this issue and enhance student engagement in testing when they write code, we propose to incentivize students to test more by gamifying continuous integration. For this we use \toolname, a tool that is seamlessly integrated into the Jenkins continuous integration platform and uses game elements based on commits to the source code repository: Developers can earn points by completing test challenges and quests generated by \toolname, compete with other developers or teams on a leaderboard, and receive achievements for their test-related accomplishments. In this paper, we present our integration of \toolname into an undergraduate-level course on software testing. We observe a correlation between how students test their code and their use of \toolname, as well as a significant improvement in the accuracy of their results compared to a previous iteration of the course without gamification. As a further indicator of how this approach improves testing behavior, the students reported enjoyment in writing tests with \toolname.
	\end{abstract}

	
	
	\maketitle
	
	\section{Introduction}
	\input{sections/introduction}

	\section{Background}
	\input{sections/background}

	\section{Gamifying Continuous Integration}
	\input{sections/gamekins}

	\section{Experiment Setup} \label{sec:setup}
	\input{sections/setup}

	\section{Results}
	\input{sections/results}

	\section{Related Work}
	\input{sections/relatedwork}

	\section{Conclusions}
	\input{sections/conclusion}

	To support replications, all source code and experiment materials used in our study are available at:
	\begin{center}
		\url{https://doi.org/10.6084/m9.figshare.23600925}
	\end{center}
	\toolname is available at:
	\begin{center}
		\href{http://gamekins.org}{https://gamekins.org}
	\end{center}
	
	\begin{acks}
		This work is supported by the DFG under grant FR 2955/2-1.
	\end{acks}

	\balance
	
	\bibliographystyle{ACM-Reference-Format}
	\bibliography{bib}
	
\end{document}

%% file: sections/introduction.tex
Software testing is a well-established concept and is extensively used in the industry~\cite{DBLP:conf/esem/SantosMCSCS17}. However, despite the availability of various tools that facilitate the testing process, such as testing support in integrated development environments (IDE)~\cite{1463097} and automated test execution in continuous integration (CI) platforms~\cite{booch1991object}, testing is often overlooked in practice. Speculated reasons for this include developers' lack of motivation to engage in testing activities and a lack of education in software testing~\cite{kapur2017release, seth2014organizational, DBLP:journals/infsof/DeakSS16, DBLP:journals/corr/WaychalC16, DBLP:journals/software/WeyukerOBP00}.
Although the often inadequate treatment of testing in higher education~\cite{seth2014organizational} has recently been countered with a growing recognition of the importance of testing~\cite{DBLP:conf/sigcse/Jones01, DBLP:conf/acse/Carrington97, DBLP:conf/iticse/MarreroS05}, this is nevertheless hampered by learners just like developers tending to perceive testing as tedious and boring~\cite{krutz2014using}.
As a result, the industry suffers significant losses due to inadequate software quality and insufficient testing~\cite{pooreport}.

Gamification techniques provide an opportunity to motivate students to test despite their perception of it being boring or tedious. Gamification involves incorporating elements commonly found in games, such as leaderboards and achievements, into non-game contexts~\cite{DBLP:conf/mindtrek/DeterdingDKN11}. It has been demonstrated that this approach helps engage students in software engineering education~\cite{Fulcini2023}, for example by gamifying lectures or tools
to teach software testing~\cite{Blanco2023, DBLP:conf/sbqs/JesusPFS19, Moreira2022, elbaum2007bug, fu2016gamification}.
However, in order for students to internalize testing and establish it as an integral part of their development approach, we argue that seamless integration of gamification of testing into the software development process is necessary.

In order to achieve this, we integrate gamification into continuous integration (CI) systems used by students. 
Specifically, we use \toolname~\cite{blinded}, a plugin designed for the widely used CI platform Jenkins\footnote{\url{https://www.jenkins.io/}}, which seamlessly integrates gamification into the software development workflow. 
By analyzing source code and test results, \toolname identifies areas where testing can be improved based on code coverage~\cite{DBLP:journals/csur/ZhuHM97} or mutation analysis~\cite{5487526} and then incentivizes improvement using gamification concepts such as challenges and quests, rewarding good testing with points, achievements, and leaderboard rankings.
%
%
We evaluated the benefits of integrating \toolname in an undergraduate software testing course, investigating its impact on the course' learning objectives and student behavior.
In detail, the contributions of this paper are as follows:
\begin{itemize}
	\item We propose the use of gamification as a means to incentivize and reward testing activities in CI.
	\item We introduce \toolname as a tool to be used in an undergraduate-level software testing course.
	\item We empirically evaluate the effects of integrating \toolname into the software testing course, comparing to a previous cohort without gamification and surveying the students.
\end{itemize}

The results of the study demonstrate a correlation between students' testing behavior and the use of \toolname, as well as a significant improvement in correct results compared to the previous edition of the course, where no gamification was used on the same assignments. While not all aspects of \toolname were universally liked by students, overall, they enjoyed its use throughout the course, confirming that \toolname is a viable teaching tool.

%% file: sections/background.tex
\subsection{Software Testing}

Software developers have access to various tools and metrics that help them identify how to improve their tests, such as code coverage, which identifies statements, branches, or execution paths that have not been covered by tests. This measurement is achieved by recording the parts of the application that are reached during test execution, resulting in lists of uncovered lines of code, percentages of covered code, and visualizations of the coverage~\cite{DBLP:journals/csur/ZhuHM97}.
Mutation analysis~\cite{5487526} assesses the effectiveness of test suites by inserting artificial faults (mutants) into a program and checking whether the tests can detect them and provides guidance on additional tests if not.
%
Static code analysis is used to identify common patterns in the code that may lead to potential bugs. For example, patterns such as code~\cite{DBLP:books/daglib/0019908} and test~\cite{van2001refactoring} smells suggest where code should be refactored or removed by developers to improve code quality.

Although there are various metrics and tools for software testing, they are often not widely used in an automated manner~\cite{DBLP:conf/sigsoft/BellerGPZ15}. Continuous integration (CI) is a common solution to this problem: CI allows developers to automatically execute their tests and calculate metrics such as code coverage after each push to a remote repository~\cite{booch1991object}. This enables developers to focus on writing code and tests in their IDE, while the CI system performs analyses and provides a summary of findings and uncovered code~\cite{spillner2019basiswissen}.
The use of CI platforms has been shown to improve productivity and enable faster software releases with higher quality~\cite{DBLP:journals/ese/ZampettiVPCGP20, DBLP:journals/ese/SoaresSSCK22}.

\subsection{Gamification of Software Testing}

The lack of motivation to write tests and to use test automation can at least partially be attributed to the perception of testing as being tedious, stressful, uncreative~\cite{DBLP:journals/corr/WaychalC16, DBLP:journals/software/WeyukerOBP00, kapur2017release}, or not appreciated~\cite{DBLP:conf/esem/SantosMCSCS17, DBLP:journals/infsof/DeakSS16}.
Motivation, which is crucial for developer productivity~\cite{DBLP:conf/esem/FrancaSS14, francca2014theory, 8370133}, can be intrinsic, describing the inner willingness to participate in an activity, or it can refer to an external signal related to a task.
Gamification is a way to provide external motivation, and it does so using game design elements within non-game contexts~\cite{DBLP:conf/mindtrek/DeterdingDKN11}.
Gamification has been shown to increase engagement and improve results compared to non-gamified development environments~\cite{DBLP:conf/sast/JesusFPF18, DBLP:journals/ese/StolSG22}. The most common gamification elements include points, badges, and leaderboards, but other elements such as challenges and achievements are also often applicable~\cite{robson2015all}.

In the context of software testing, gamification has been applied in both professional (e.g.,~\cite{DBLP:conf/sera/Parizi16, DBLP:conf/icse/ScherrEH18}) and educational (e.g.,~\cite{bell2011secret, fraser2019gamifying, fu2016gamification, DBLP:conf/icsob/Yordanova19}) settings. Previous attempts to gamify software testing education have focused on specific aspects such as mutation testing~\cite{fraser2019gamifying} or security testing~\cite{fu2016gamification}. Our aim is orthogonal: Rather than teaching specific test techniques, we aim to get students to integrate testing into their regular software development routine.


%% file: sections/gamekins.tex
Despite its importance, testing is not tightly integrated into most computer science curricula. Very often, it is part of general software engineering courses, where time dedicated to testing tends to be limited~\cite{GAROUSI2020110570}. However, dedicated software testing courses are becoming more common. Dedicating an entire course to software testing provides the opportunity to cover many different facets of testing, ranging from testing techniques to automating testing, and also how to firmly integrate testing during software development.

\subsection{Challenges in Teaching Software Testing}

At the \university, we established a mandatory undergraduate course on software testing in 2018. Grading is based entirely on coursework, forcing the students to engage practically with testing challenges at different levels. Besides covering the standard portfolio of techniques common in software quality assurance, where testing in practice is often conducted by dedicated software testers, we also aim to cover \emph{developer testing}, i.e., the tight integration of testing into regular software development activities. Thus, one of the course aims is to establish testing as a routine activity that students perform when developing their own software.

However, throughout multiple iterations of the course, we observed several issues inhibiting this aim: Students exhibit a lack of motivation to write tests, which results in them writing only the bare minimum number of tests required to meet a certain goal (e.g., if code coverage is required). This also indicates that students tend to overly focus on metrics, in particular code coverage, which is easy to obtain and understand.
Further evidence for the lack of engagement can be found in the students' commit logs: Rather than including tests throughout the process and using fine-grained commits~\cite{singer2012bit}, testing is usually conducted as a post-hoc exercise. Anecdotally, we also observed general weaknesses such as hard-coded test data (e.g., paths to test resources) which may not be reflected in the metrics students aim for but suggest a lack of proper engagement with testing. Consequently, there is a need for further incentives to establish testing in regular software development.

\subsection{Gamification Elements of \toolname}

We aim to improve testing courses by incentivizing students to test more and better using the CI system underlying their coding tasks. For this, we developed a tool, \toolname~\cite{blinded}\footnote{Available at \href{http://gamekins.org}{https://gamekins.org}}, which incorporates gamification elements into CI based on static code, coverage, and mutation analysis. The combination of underlying quality metrics is intended to shift the students' focus away from solely relying on one metric.
To enforce better commit and testing behavior, \toolname provides gamification elements only after new commits.

The general workflow of \toolname is as follows\footnote{\label{not:toolpaper}Detailed information can be found in \cite{blinded}}: Developers commit code changes to a version control system, which triggers a job in the Jenkins CI environment. Jenkins executes a build job, which includes running the project's tests, and then invokes \toolname. \toolname utilizes information from the run and the version control system to update and generate challenges, quests, achievements, and the leaderboard. 

\subsubsection{Challenges}

\begin{figure*}[tb]
	\includegraphics[width=0.9\textwidth]{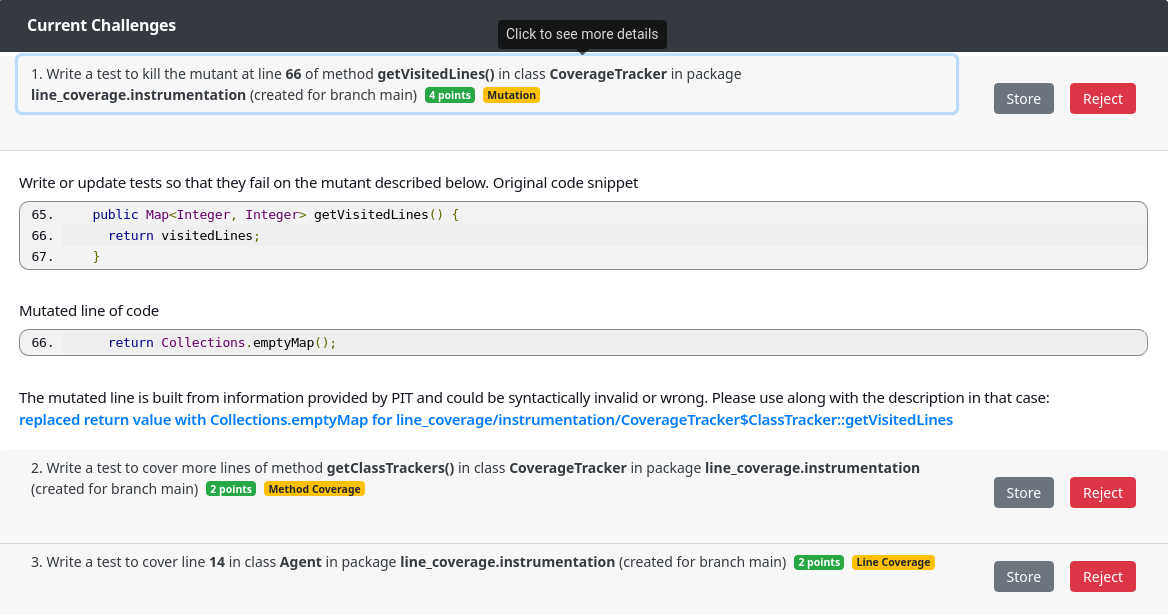}
	\caption{This overview highlights the current challenges, including a Line Coverage, a Method Coverage, and a Mutation Challenge. The Mutation Challenge is further explained with a code snippet, the mutated line of code, and a description.}
	\label{fig:gamekinschallenges}
\end{figure*}

\toolname offers a variety of seven distinct challenge types, each designed to provide developers with specific test- and quality-related tasks to complete:\footnoteref{not:toolpaper}
\begin{itemize}
	\item \textbf{Build Challenge}: This challenge requires developers to resolve a build failure. To prevent misuse of \toolname by repeatedly breaking and fixing the build for points, this challenge is generated only once per week.
	\item \textbf{Test Challenge}: This generic challenge requires developers to write at least one additional test without specifying the specific code area to target.
	\item \textbf{Class Coverage Challenge}: This challenge type focuses on increasing code coverage for a selected Java class.
	\item \textbf{Method Coverage Challenge}: This challenge focuses on improving the coverage of a specific target method.
	\item \textbf{Line Coverage Challenge}: This challenge focuses on improving the coverage of a specific line of code that is not fully covered. The task is to either cover or increase the branch coverage of the line.
	\item \textbf{Mutation Challenge}: This challenge focuses on detecting a specific mutant by adding or improving a test, and therefore goes beyond a simple check of the robustness of the test suite to closing test gaps~\cite{DBLP:journals/tse/PetrovicIFJ22}. The mutants are generated with PIT\footnote{\url{https://pitest.org/}}, a popular mutation testing tool for Java.
	\item \textbf{Smell Challenge}: This challenge involves analyzing a target class using SonarLint\footnote{\url{https://www.sonarlint.org/}}, a code quality tool. The challenge is to choose one of the detected smells in either the source or test files and remove it.
\end{itemize}
In order to generate new challenges, \toolname ensures that there is always a (configurable) fixed number of current challenges. When generating a new challenge, \toolname only considers recently changed classes based on the commit history. This ensures that the challenge is relevant to the functionality currently considered by the developer. To generate a challenge, \toolname first ranks all classes in the project under test based on the current code coverage, then probabilistically selects a class for the challenge, giving higher chances to classes with lower coverage; finally, a challenge type is randomly selected, and a random instance for this type is generated. 

If a developer already has open challenges, \toolname checks if these challenges have been solved. If they have been solved, \toolname will generate new challenges to replace them. Additionally, \toolname also checks if all currently open challenges are still applicable, for example by verifying if the specific code fragment for a challenge has not been deleted. To keep developers informed, all current challenges are displayed on a dedicated page in Jenkins, which provides an overview of open challenges and their status (\cref{fig:gamekinschallenges}). Clicking on a challenge provides additional information, including the code snippet mentioned in the challenge description, as well as an explanation of what needs to be done. This information is particularly important for Mutation Challenges, where both the modified source code and the original code are shown.

To remove a current challenge from the list, participants can click on the \texttt{Reject} button. When rejecting a challenge, participants are required to provide a reason, which aims to limit rejections and helps to improve \toolname. Some possible reasons for rejecting a challenge include cases where the challenge is unnecessary due to defensive programming or when reported code smells are intentionally present in the code.
In some cases, challenges may be automatically rejected if, for example, the target line or class has been removed. If a Class Coverage Challenge is rejected, this will block the future generation of any challenges in the class that the rejected challenge was targeting. However, this can be undone in the list of rejected challenges if necessary.

\subsubsection{Quests}

\begin{figure}[tb]
	\includegraphics[width=\linewidth]{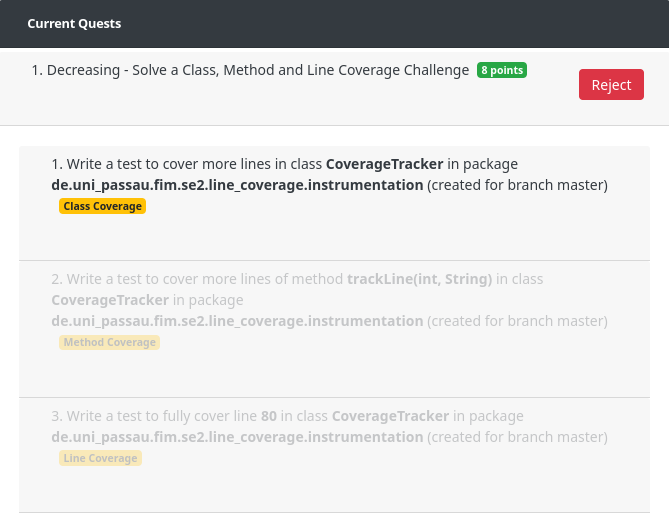}
	\caption{Example quest consisting of three Smell Challenges, with the active one highlighted}
	\label{fig:gamekinsquests}
\end{figure}

Quests are a way to group multiple challenges together. They consist of a series of individual challenges that need to be solved one after the other. In a quest, only the current challenge that needs to be solved next is enabled and can be expanded to show more details (\cref{fig:gamekinsquests}). The successive challenges in the quest are disabled and cannot be viewed until it is their turn to be solved. To earn points for a quest, all steps must be solved in succession.
\toolname supports a total of nine different types of quests\footnoteref{not:toolpaper}.

Quests are created by checking the prerequisites of potential quests, such as ensuring that a sufficient number of required challenges can be generated. One quest is then randomly chosen from the available options. The individual steps or challenges within a quest are generated as if they were standalone challenges but with predetermined classes based on the quest type. Participants have the option to reject quests, or quests can be automatically rejected by \toolname if at least one of the challenges within the quest becomes unsolvable. In this case, participants will receive points for the challenges they have already solved within the quest, but not for the steps themselves.

\subsubsection{Leaderboard}

\begin{figure}[tb]
	\includegraphics[width=\linewidth]{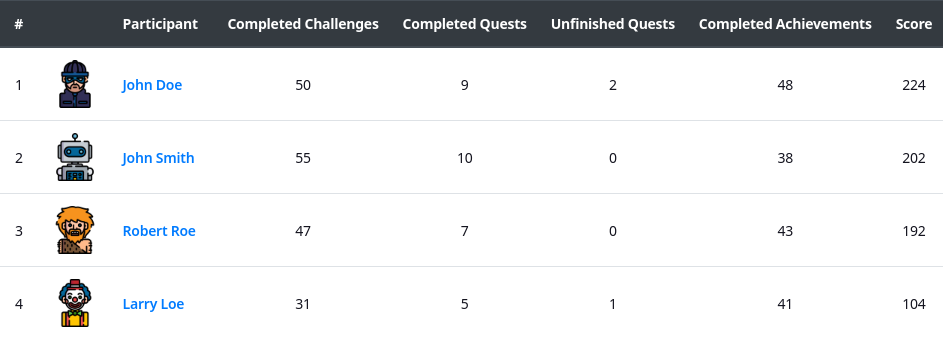}
	\caption{The project leaderboard displays the completed challenges, as well as (unfinished) quests and achievements. It also shows the score and avatar of each user. Clicking on their name will take you to their achievements..}
	\label{fig:gamekinsleaderboard}
\end{figure}

By completing challenges and quests, users earn points, and their point rankings are displayed on a leaderboard in Jenkins. The leaderboard not only shows the points but also the total number of completed challenges and achievements, fostering competition.
Participants can personalize their experience by selecting one of the 50 avatars to be displayed on the leaderboard.

\subsubsection{Achievements}

\begin{figure*}[tb]
	\hspace*{\fill}
	\begin{subfigure}[t]{0.35\textwidth}
		
		\includegraphics[width=\textwidth]{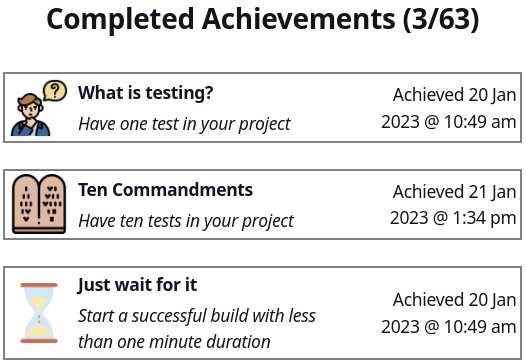}
		
		\caption{An example list of completed achievements}
		\label{fig:gamekinsachievementscompleted}
	\end{subfigure}
	\hfill
	\begin{subfigure}[t]{0.35\textwidth}
		
		\includegraphics[width=\textwidth]{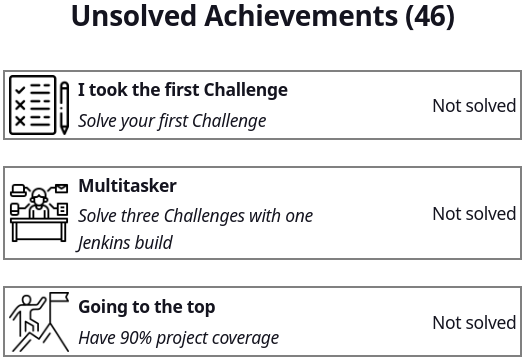}
		
		\caption{An example list of unsolved achievements}
		\label{fig:gamekinsachievementsunsolved}
	\end{subfigure}
	\hspace*{\fill}
	
	\caption{The list of achievements in \toolname is categorized into completed, unsolved, and secret ones. Each achievement has a title, a description, a date when it was solved, and an icon. Once solved, its icon is colored to indicate completion.}
	\label{fig:gamekinsachievements}
\end{figure*}

Developers are rewarded for their test achievements, which are based on specific behaviors or actions they perform, regardless of specific challenges. These achievements have varying difficulty, ranging from easy tasks, like having a test in a project, to more challenging ones, like achieving 100\,\% coverage. Currently, \toolname implements 63 achievements, and developers can view obtained and open achievements (\cref{fig:gamekinsachievements}). Some achievements are kept secret and are only revealed once they are completed.

\subsection{Gamified Elements and the Testing Curriculum}

Besides offering general incentives, the integration of \toolname into coursework allows to address several learning objectives outlined in the \emph{Curriculum Guidelines for Undergraduate Degree Programs in Software Engineering}~\cite{7328640}, such as:
\begin{itemize}
	\item Static analysis (VAV.rev.3) and refactoring (PRO.evo.3), e.g., with Smell Challenges 
	\item Unit (VAV.tst.1) and integration (VAV.tst.5) testing with Challenges and Quests
	\item Testing edge cases and boundary conditions with Mutation Challenges (VAV.tst.2)
	\item Coverage analysis with Coverage Challenges (VAV.tst.3)
	\item Regression testing (VAV.tst.10), testing tools and automation (VAV.tst.11) with \toolname as CI plugin
	\item Build processes (PRO.cm.4) and analyzing failure reports (VAV.par.1) with \toolname as CI plugin
\end{itemize}
%

%% file: sections/setup.tex
In order to evaluate whether \toolname influences the testing behavior of students, we integrated it into our software testing course and aim to answer the following questions:
\begin{itemize}
	\item \textbf{RQ1}: How did the students use \toolname during the course?
	\item \textbf{RQ2}: What testing behavior did the students exhibit?
	\item \textbf{RQ3}: How did the students perceive the integration of \\\toolname into their projects?
\end{itemize}

\subsection{Software Testing Course}

During the computer science Bachelor program at the \university, students are required to take a course on Software Testing. This course includes two hours of lectures and two hours of exercise every week. The coursework consists of five coding projects that students have to implement throughout the course:
\begin{itemize}
	\item \textbf{Introduction}: The initial assignment requires students to create and test a small class using JUnit5\footnote{\url{https://junit.org/junit5/}}.
	\item \textbf{Test-driven development}: For the second assignment, students are tasked with implementing the model of a canteen web application to display the current menu. The goal is to create it in a test-driven manner, where students have to write their tests before implementing any features~\cite{DBLP:books/daglib/0019907}.
	\item \textbf{Behavior-driven development}: In the third task of the assignment, students are required to create the view for the already implemented model of the canteen web app. To test the graphical user interface (GUI), students have to utilize Selenium\footnote{\url{https://www.selenium.dev/}}, a web testing framework, and Cucumber\footnote{\url{https://cucumber.io/}}, a tool for behavior-driven development.
	\item \textbf{Line-coverage analyzer (Analyzer)}: In the fourth project, students are required to implement a simple line-coverage analyzer. This analyzer will track the lines of code that are visited during the execution of tests.
	\item \textbf{Coverage-based fuzzer (Fuzzer)}: In the last assignment, students have to implement a coverage-based fuzzer, which 
	uses coverage information to guide the generation of inputs and to prioritize the exploration of untested code paths.
\end{itemize}

The students' grades are determined based on the performance and correctness of the five projects. The main objective is to implement the required functionality for each project and thoroughly test it with 100\,\% code coverage. Only the last two projects (Analyzer and Fuzzer) were used for the evaluation.

\subsection{Integration of \toolname} \label{sec:integration}

The integration of \toolname focuses on the last two tasks, as the introduction task is too small to viably use CI, and for the test-driven and behavior-driven tasks we require the commit history to show evidence of a correct implementation of these approaches, where \toolname would interfere.
As one of the requirements for the last two assignments, the students are mandated to use \toolname by writing tests based on the challenges generated by \toolname. Since students are evaluated individually, they cannot utilize the team functionalities of \toolname. 
The workflow for the students is as follows: They select a challenge to tackle, write the corresponding test in their IDE, and commit and push the test to allow \toolname to verify it. The students' grades are based on their adherence to this workflow, and they can conclude their testing efforts when \toolname is no longer able to generate new challenges.

\subsection{Participants}

The participants of our study are the students enrolled in the Software Testing course in the winter semester of 2022/23. There were a total of 26 students who completed the Analyzer project and 25 students for the Fuzzer project (one student dropped out after completing the Analyzer). We retrieved their consent for anonymized usage of data collected by \toolname.
Out of the total participants, we received survey responses from 17 students, providing valuable insights for our evaluation (as discussed in \cref{sec:evalRQ4}). Among the respondents, two out of 17 students identify as female.
Participants were primarily in their beginning and mid-twenties, and the majority of students (65\,\%) claimed between one and three years of experience with Java. Additionally, most students (35\,\%) claimed between three and six months of experience with JUnit, followed by 29\,\% of students with six to twelve months. 

\subsection{Data Analysis}

The analysis of the experiment involves comparing the results obtained from the participating students. In order to determine the significance of any differences, we employ the exact Wilcoxon-Mann-Whitney test~\cite{10.1214/aoms/1177730491} with a significance level $\alpha = 0.05$.
%

\subsubsection{RQ1: How did the students use \toolname during the course?}

The data collected by \toolname, including current, completed, and rejected challenges and quests, are stored in the configuration files of each user. This data can be easily extracted for further evaluation, which is the focus of our analysis in this research question.
We consider the differences between the Analyzer and Fuzzer projects to determine if the students performed or used \toolname differently during these projects. We examine the various types of challenges solved by the students and investigate the reasons why they rejected certain challenges. This analysis helps us identify any difficulties that the students faced while using \toolname.

\subsubsection{RQ2: What testing behavior did the students exhibit?}
We compare both projects in terms of (1) number of tests, (2) number of commits, (3) line coverage, (4) mutation score, and (5) grade. The line coverage of the participants' implementations is measured with the help of JaCoCo\footnote{\url{https://www.jacoco.org/jacoco/}}, while we use PIT to determine the mutation score. Coverage is a common metric to determine how well the source code of a project is exercised~\cite{ivankovic2019code}, while mutation analysis detects test gaps also within covered code~\cite{DBLP:journals/tse/PetrovicIFJ22}.
The grade is an individual score assigned to each participant for each project, ranging from 0 to 100. It is determined using several factors, including average line and branch coverage, mutation score, and the usage of \toolname. The usage of \toolname depends on the participants' engagement in writing tests and solving challenges using the tool (see \cref{sec:integration}). For the Analyzer project, the correct output 
is also considered in the grading process by running the participants' implementation against eleven different examples and comparing their output with a reference implementation.
To further analyze the relationship between the participants' performance in \toolname and their project outcomes, we use Pearson correlation~\cite{cohen2009pearson} to determine if there is a correlation between the scores achieved by the participants in \toolname and the number of tests, number of commits, code coverage, mutation score, and grade.

The Analyzer project was previously used in the course in 2019, with the same task and framework, differing only in the use of \toolname. We compare the results of 2019 and 2022 in terms of (1)~number of tests, (2) number of commits, (3) line coverage, (4) mutation score, and (5) correct output. For this, we count the number of tests in both the 2019 and 2022 projects and execute them to obtain coverage using JaCoCo. Next, we calculate mutation scores for the student code using PIT. Lastly, we compare the output generated by their code with the output of our reference solution.

\subsubsection{RQ3: How did the students perceive the integration of \toolname into their projects?} \label{sec:evalRQ4}

\begin{table}[t]
	\centering
	\caption{Survey questions \\ \tiny{with Single Choice as SC and Multiple Choice as MC}}
	\label{tab:allquestions}
        \vspace{-1em}
	\resizebox{\columnwidth}{!}{
		\begin{tabular}{lp{7cm}l}
			\toprule
			ID & Question & Type     \\ \midrule
			\addlinespace[0.5em]
			\multicolumn{2}{l}{Questions in the category participant demographics} \\ \cmidrule(r){1-2}
			P1 & Age                     & Free-text   \\ 
			P2 & Gender       & SC + free-text  \\ 
			P3 & Experience with Java         & SC   \\ 
			P4 & Experience with JUnit                   & SC   \\ 
			\addlinespace[0.5em]
			\multicolumn{2}{l}{Questions in the category \toolname} \\ \cmidrule(r){1-2}
			G1 & I enjoyed using \toolname                     		& Likert 5 points   	\\ 
			G2 & I enjoy writing unit tests even when it is not part of a game                     		& Likert 5 points   	\\ 
			G3 & Writing unit tests as part of the game is more fun than writing unit tests while coding                     		& Likert 5 points   	\\ 
			G4 & I learned/practised useful skills using \toolname                     		& Likert 5 points   	\\ 
			G5 & I would consider using \toolname in other projects                     		& Likert 5 points   	\\ 
			G6 & I liked this part of \toolname -- Challenges      		& Likert 5 points   	\\ 
			G7 & I liked this part of \toolname -- Quests      		& Likert 5 points   	\\ 
			G8 & I liked this part of \toolname -- Achievements      		& Likert 5 points   	\\ 
			G9 & I liked this part of \toolname -- Leaderboard      		& Likert 5 points   	\\ 
			G10 & I liked this part of \toolname -- Avatars      		& Likert 5 points   	\\ 
			G11 & Is there a special reason why you liked or disliked a part of \toolname?        						& Free-text   	\\ 
			G12 & Did you ever feel stressed or pressured to solve challenges by \toolname or external factors?    & Yes/No + free-text   \\ 
			G13 & Did you write more tests while using \toolname than you would have otherwise?					& SC 	\\
			G14 & Did you write unnecessary tests to solve challenges? 					& Yes/No 	\\
			G15 & Did you find a bug using tests written for \toolname?					& Yes/No 	\\
			\addlinespace[0.5em]
			\multicolumn{2}{l}{Questions in the category further thoughts} \\ \cmidrule(r){1-2}
			F1 & What did you like about \toolname?               	& Free-text   	\\ 
			F2 & What did you not like about \toolname?              	& Free-text   	\\ 
			F3 & Do you have a suggestion for improvement?              	& Free-text   	\\ 
			F4 & Is there something else you want to mention?               	& Free-text   	\\ 
			\bottomrule
		\end{tabular}%
	}
\end{table}

To address this research question, we administered a survey to the students, comprising 23 questions categorized into three sections (\cref{tab:allquestions}). The initial section focused on gathering demographic information about the participants, while the second section inquired about their experience with \toolname. The final section allowed students to provide feedback on their likes and dislikes regarding \toolname. We present the survey responses using Likert plots and analyze the students' free-text answers to gain insights into their perceptions of \toolname.

\subsection{Threats to Validity}

\emph{Threats to external validity} may arise due to the limited number of participants, which  limits generalizability. It is important to note that our software testing course is a mandatory course at the \university, meaning that every student in their Bachelor's program is required to take it. This could potentially impact the behavior and performance of the students compared to those in non-mandatory courses. Additionally, the course is specifically designed for students in their fifth semester who possess a certain level of computer science knowledge. Students in different semesters or with varying levels of knowledge may exhibit different behaviors and outcomes when using \toolname. Moreover, students from different universities, countries, or degree programs may also exhibit different behaviors and outcomes.

\emph{Threats to internal validity} could potentially arise from errors in our data collection infrastructure, plugin, and the integration of \toolname into the course. However, we took measures to mitigate these threats by thoroughly testing \toolname and immediately addressing any issues arising during the course. Another potential threat to internal validity is the possibility of differences in knowledge levels between the students in 2019 and 2022. To minimize this threat, we conducted exercise sessions during the projects where all students had the opportunity to ask questions and clarify any uncertainties regarding the tasks. Additionally, there is a possibility that students may have collaborated instead of completing the tasks individually. However, we conducted a plagiarism check on their code and did not find any irregularities.

\emph{Threats to construct validity} may arise due to the mandatory use of \toolname and the grading associated with it, which could potentially influence their behavior and perception of the tool.

%% file: sections/results.tex
\subsection{RQ1: How did the students use \toolname during the course?}

\begin{figure*}
	\centering
	\begin{subfigure}[t]{0.245\textwidth}
		\centering
		\includegraphics[width=\textwidth]{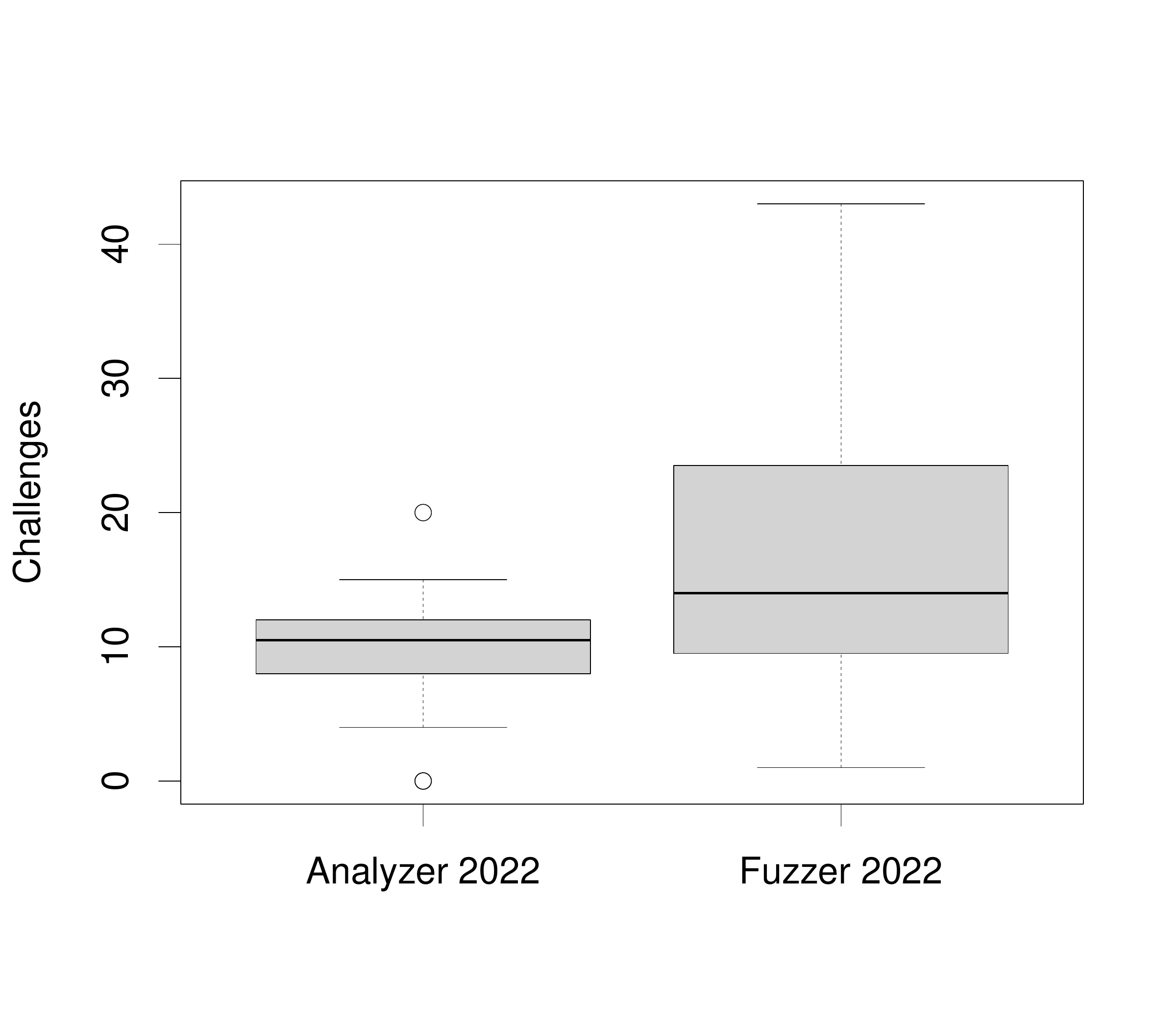}
		\vspace{-2em}
		\caption{Number of challenges}
		\label{fig:challenges}
	\end{subfigure}
	\hfill
	\begin{subfigure}[t]{0.245\textwidth}
		\centering
		\includegraphics[width=\textwidth]{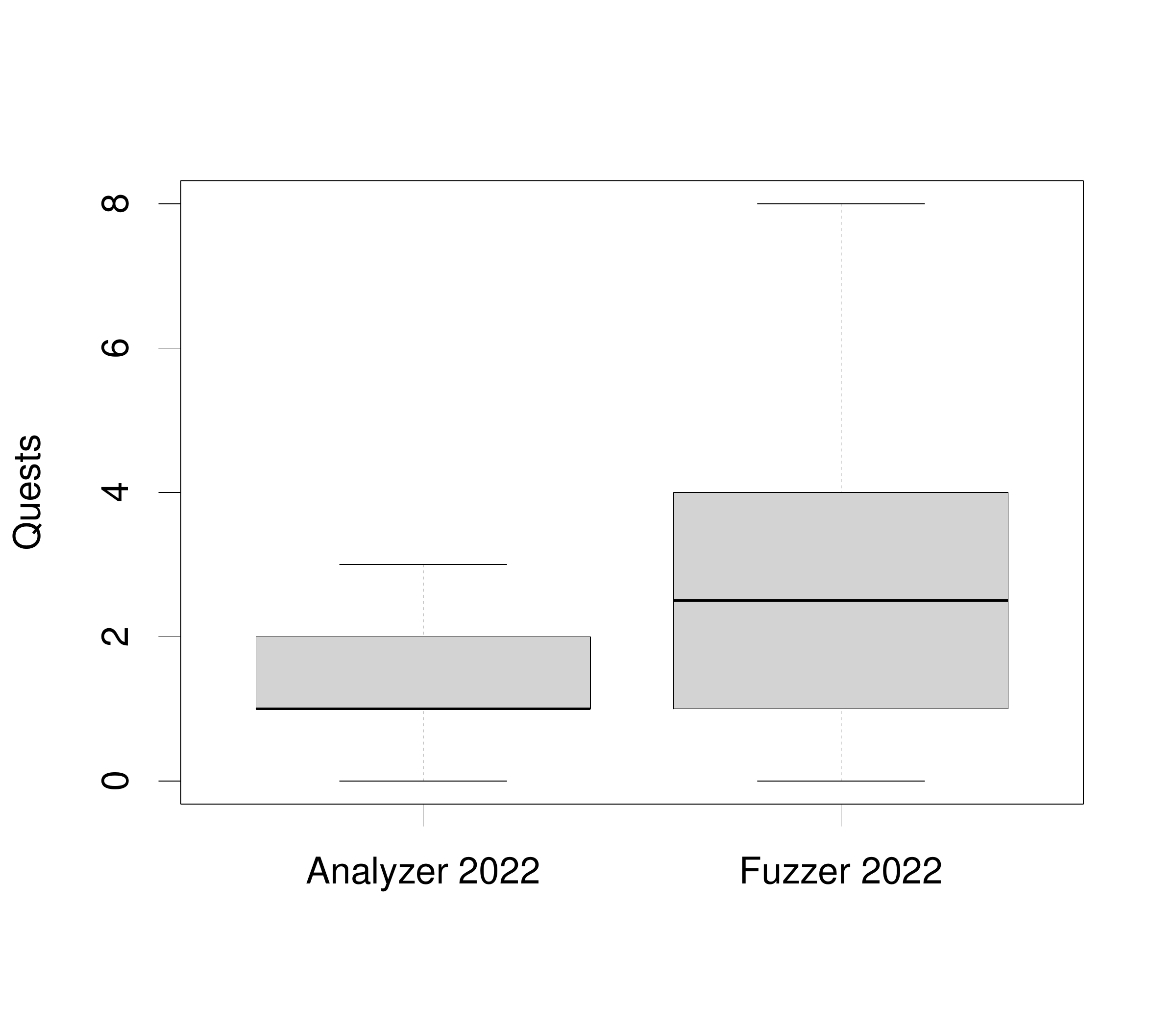}
		\vspace{-2em}
		\caption{Number of quests}
		\label{fig:quests}
	\end{subfigure}
	\hfill
	\begin{subfigure}[t]{0.245\textwidth}
		\centering
		\includegraphics[width=\textwidth]{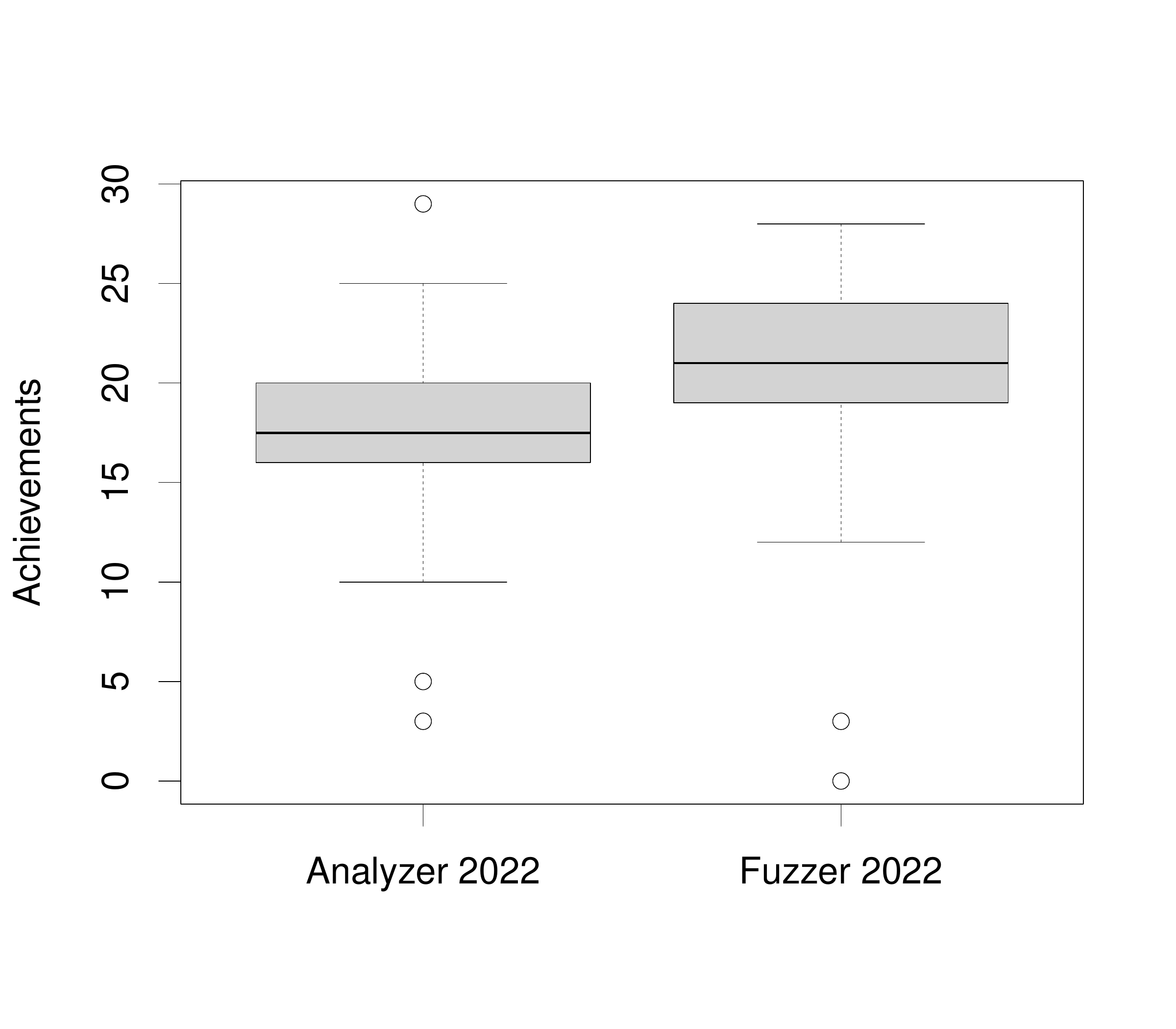}
		\vspace{-2em}
		\caption{Number of achievements}
		\label{fig:achievements}
	\end{subfigure}
	\hfill
	\begin{subfigure}[t]{0.245\textwidth}
		\centering
		\includegraphics[width=\textwidth]{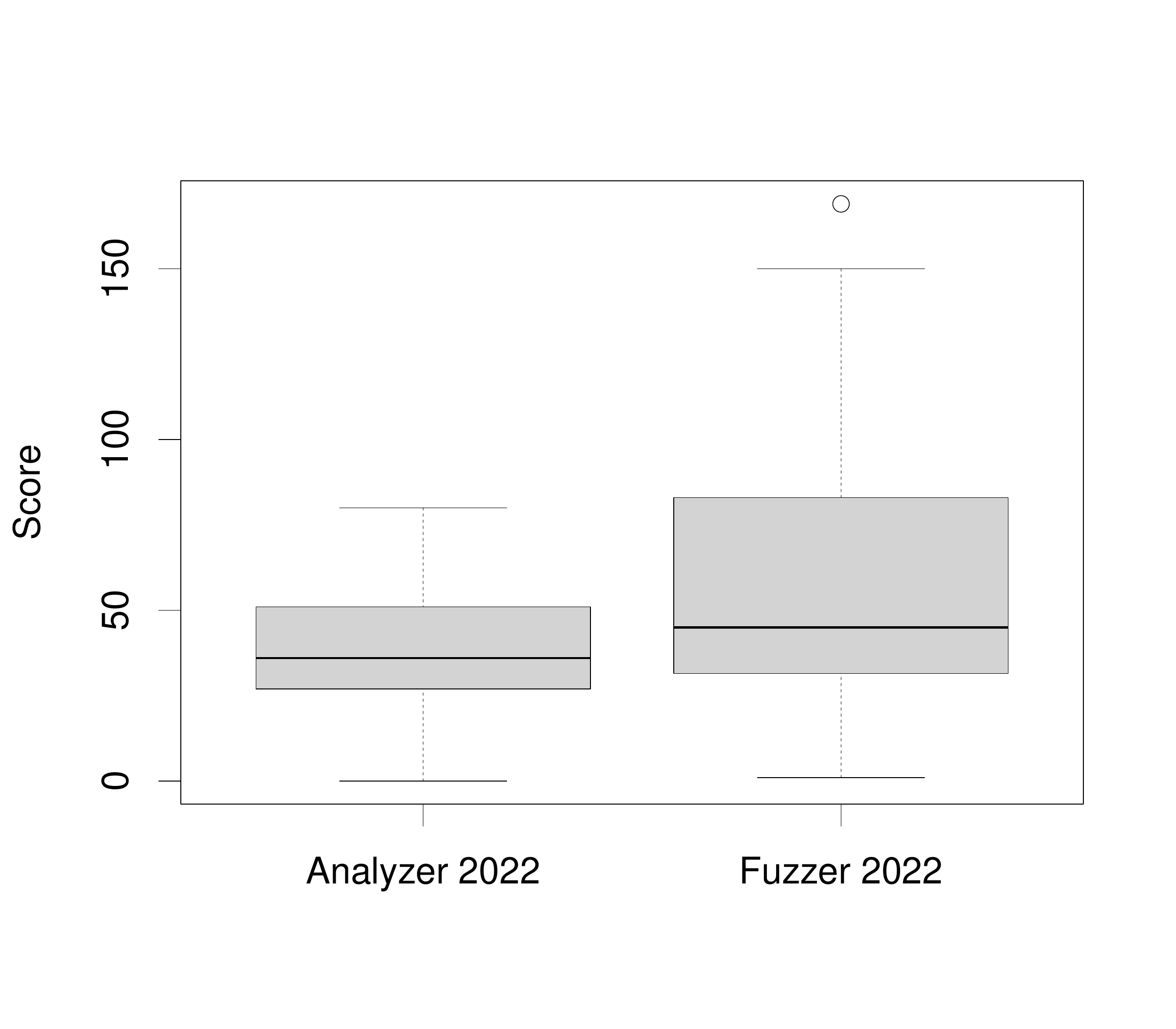}
		\vspace{-2em}
		\caption{Points in \toolname}
		\label{fig:score}
	\end{subfigure}
	
	\caption{Statistics on the use of \toolname in both projects}
	\label{fig:boxplots}
\end{figure*}

\begin{table}[]
	\caption{Total number and ratio of completed and rejected challenges}
	\label{tab:challenges}
        \vspace{-1em}
	\resizebox{\columnwidth}{!}{
		\begin{tabular}{@{}lrrrrr@{}}
			\toprule
			\multicolumn{1}{c}{{Type}} & \multicolumn{2}{c}{Completed}                          & \multicolumn{2}{c}{Rejected}                           & \multicolumn{1}{c}{{Total}} \\ \cmidrule(lr){2-5}
			\multicolumn{1}{c}{}                      & \multicolumn{1}{l}{Number} & \multicolumn{1}{l}{Ratio} & \multicolumn{1}{l}{Number} & \multicolumn{1}{l}{Ratio} & \multicolumn{1}{c}{}                       \\ \midrule
			Build Challenge                           & 32                         & 100\,\%                   & 0                          & 0\,\%                     & 32                                         \\
			Class Coverage Challenge                  & 65                         & 88\,\%                    & 9                          & 12\,\%                    & 74                                         \\
			Line Coverage Challenge                   & 123                        & 78\,\%                    & 34                         & 22\,\%                    & 157                                        \\
			Method Coverage Challenge                 & 76                         & 83\,\%                    & 16                         & 17\,\%                    & 92                                         \\
			Mutation Challenge                        & 218                        & 75\,\%                    & 71                         & 25\,\%                    & 289                                        \\
			Smell Challenge                           & 76                         & 80\,\%                    & 19                         & 20\,\%                    & 95                                         \\
			Test Challenge                            & 140                        & 97\,\%                    & 5                          & 3\,\%                     & 145                                        \\ \bottomrule
		\end{tabular}
	}
\end{table}

The students completed a total of 730 challenges, 113 quests, and 941 achievements in \toolname. They collected an average of 39.5 and 61.5 points for the Analyzer and Fuzzer projects, respectively, with \toolname (\cref{fig:score}). In the Analyzer project, on average they completed 10.2 challenges (\cref{fig:challenges}) and 1.6 quests (\cref{fig:quests}), and 17.0 challenges and 3.1 quests for the Fuzzer project. The students were new to \toolname in the Analyzer project, so they likely required more time and assistance to become familiar with it. As a result, they completed significantly more challenges ($p=0.026$) and quests ($p=0.011$) in the Fuzzer project. The number of achievements (\cref{fig:achievements}) is higher compared to challenges and quests because the achievements in \toolname target large industrial projects and may currently be configured to be too easy to obtain on the relatively small student projects. Over half (52\,\%) of the students used avatars to represent themselves in the leaderboard, indicating that they actually consulted the leaderboard. We also displayed the leaderboard during the exercise sessions to enhance visibility and competition among the students.

According to the data in \cref{tab:challenges}, the majority of the challenges solved by the students were coverage-based challenges (36.2\,\%), followed by Mutation (29.9\,\%), Test (19.2\,\%), and Smell Challenges (10.4\,\%). This distribution is expected because the coverage-based challenges are fast and easy to solve while the Mutation Challenges reveal untested code if the coverage is already high.
%
The students rejected a total of 154 challenges, with the majority of rejections occurring in the Fuzzer project (125). The reasons varied: 
\begin{itemize}
	\item Automatically rejected by \toolname (49)
	\item No idea how to test (33)
	\item Line already covered (generated because of missing branches)~(28)
	\item Not feasible to test (defensive programming) (18)
	\item Changed/deleted code (10)
	\item No mutated line available (byte code to source code translation is not always possible) (6)
	\item Mutant already killed (PIT does not recognize the mutant as killed) (6)
	\item Challenge is out of scope (generated for a part of the program the students do not need to test) (4)
\end{itemize}

\begin{figure*}
	\centering
	\begin{subfigure}[t]{0.49\textwidth}
		\centering
		\includegraphics[width=\textwidth]{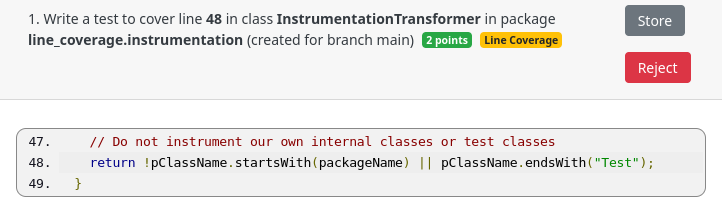}
		
		\caption{A Line Coverage Challenge that is easy to solve}
		\label{fig:challengeeasy}
	\end{subfigure}
	\hfill
	\begin{subfigure}[t]{0.49\textwidth}
		\centering
		\includegraphics[width=\textwidth]{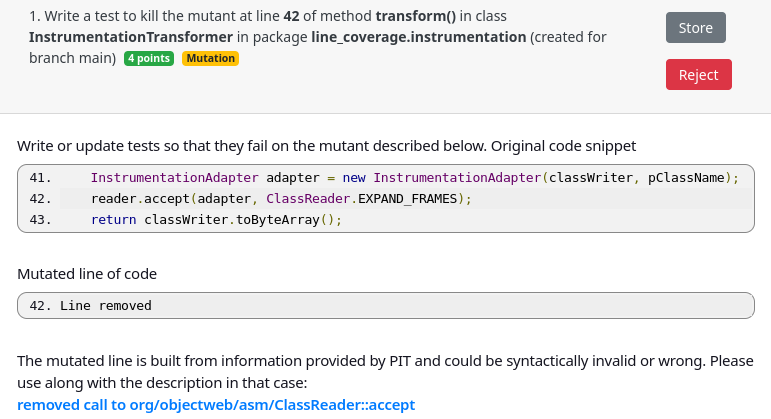}
		
		\caption{A Mutation Challenge that is hard to solve}
		\label{fig:challengehard}
	\end{subfigure}
	
	\caption{Examples for easy and hard challenges}
	\label{fig:easyhardchallenge}
\end{figure*}

The rejection rates also varied for different types of challenges. The easiest challenges, such as Build (0\,\%), Test (3\,\%), and Class Coverage Challenges (12\,\%), had the lowest rejection rates (\cref{tab:challenges}).
The most successfully solved challenge overall was the Line Coverage Challenge shown in \cref{fig:challengeeasy}. This example challenge for the Analyzer project involves checking if the class to be instrumented was not in the wrong package and not a test class and the challenge is solved by any test that calls this method.
%
%
Mutation Challenges were rejected most often overall (25\,\%), followed by Line Coverage challenges (22\,\%).
These challenges are quite specific and require testing specific lines or mutants and do not give students a chance to pick easier targets for their tests. In particular writing tests that can detect mutants can be non-trivial. For instance, \cref{fig:challengehard} represents a challenging mutant that removes a line necessary for parts of the instrumentation. Detecting this mutant in the Analyzer project requires examining the resulting byte code after instrumentation. Many students struggled with challenges like this, particularly for the Analyzer project, which required mocking, and we also had to fix some issues with PIT. 

\summary{RQ1}{Students actively used \toolname for completing challenges, quests, achievements, and checking the leaderboard. However, they often rejected challenges to retrieve easier ones.}

\subsection{RQ2: What testing behavior did the students exhibit?}

\begin{figure}
	\centering
	\begin{subfigure}[t]{0.49\linewidth}
		\centering
		\includegraphics[width=\textwidth]{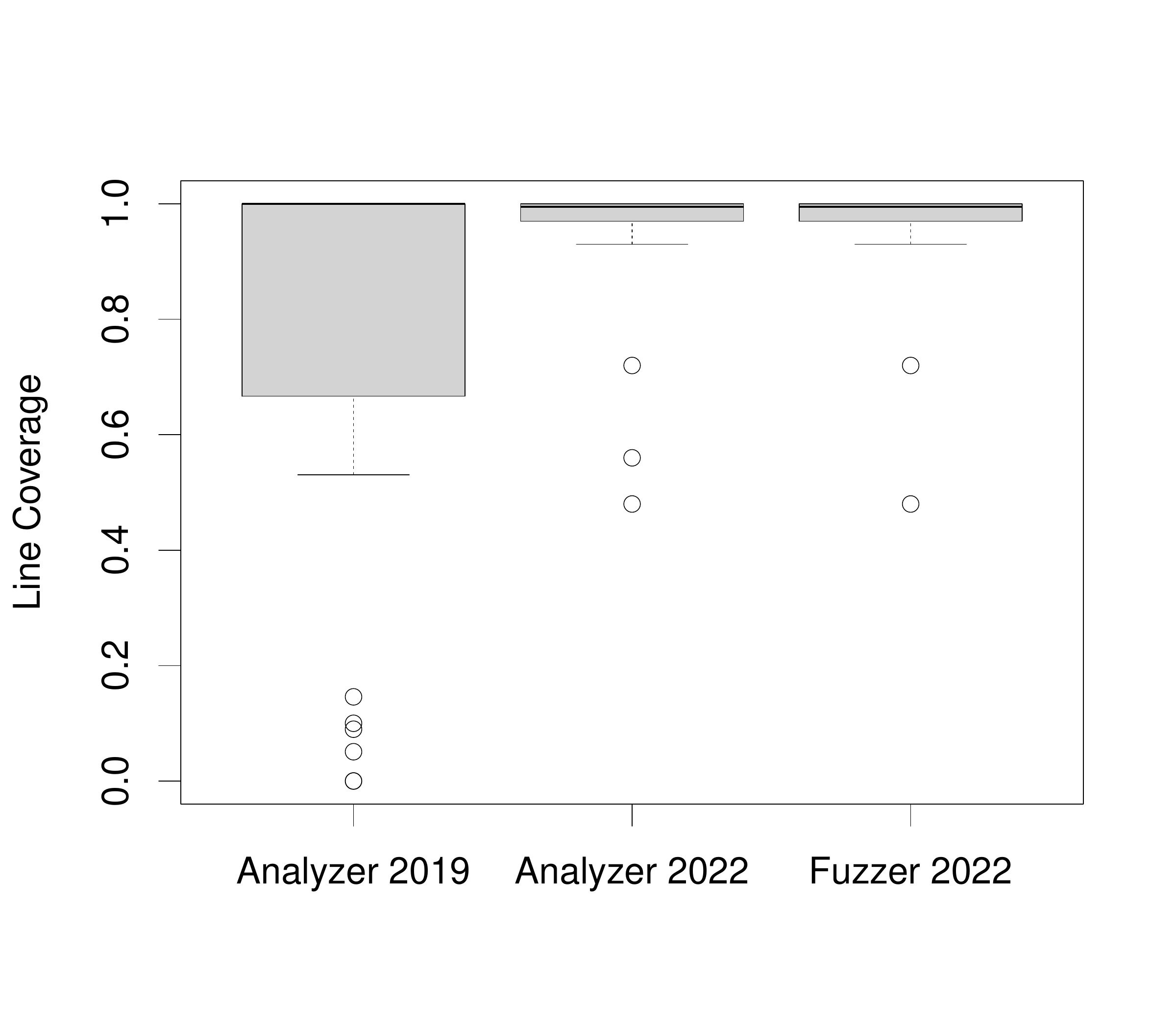}
		\vspace{-2em}
		\caption{Line coverage}
		\label{fig:coverage}
	\end{subfigure}
	\hfill
	\begin{subfigure}[t]{0.49\linewidth}
		\centering
		\includegraphics[width=\textwidth]{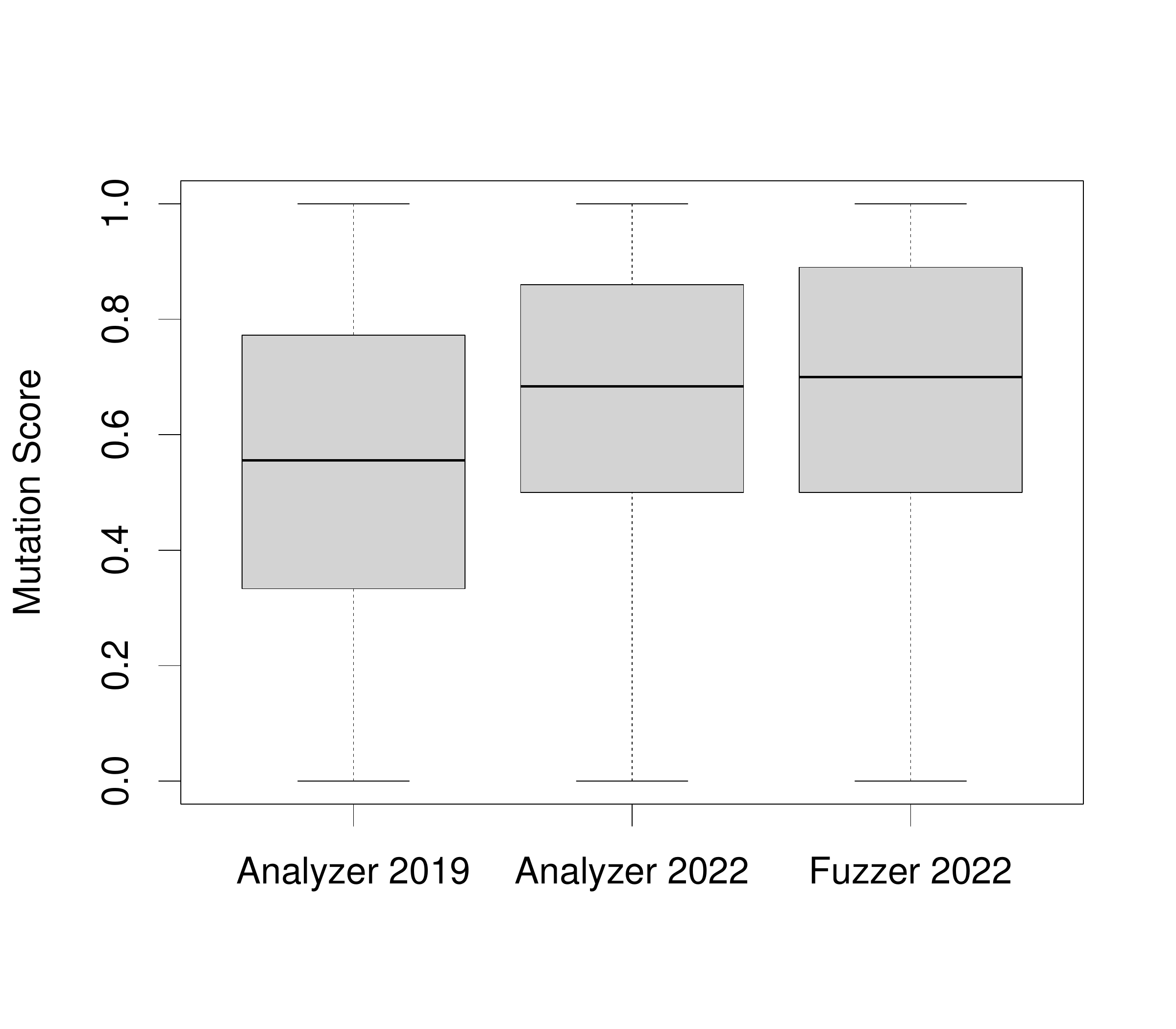}
		\vspace{-2em}
		\caption{Mutation score}
		\label{fig:mutation}
	\end{subfigure}
	\hfill
	\begin{subfigure}[t]{0.49\linewidth}
		\centering
		\includegraphics[width=\textwidth]{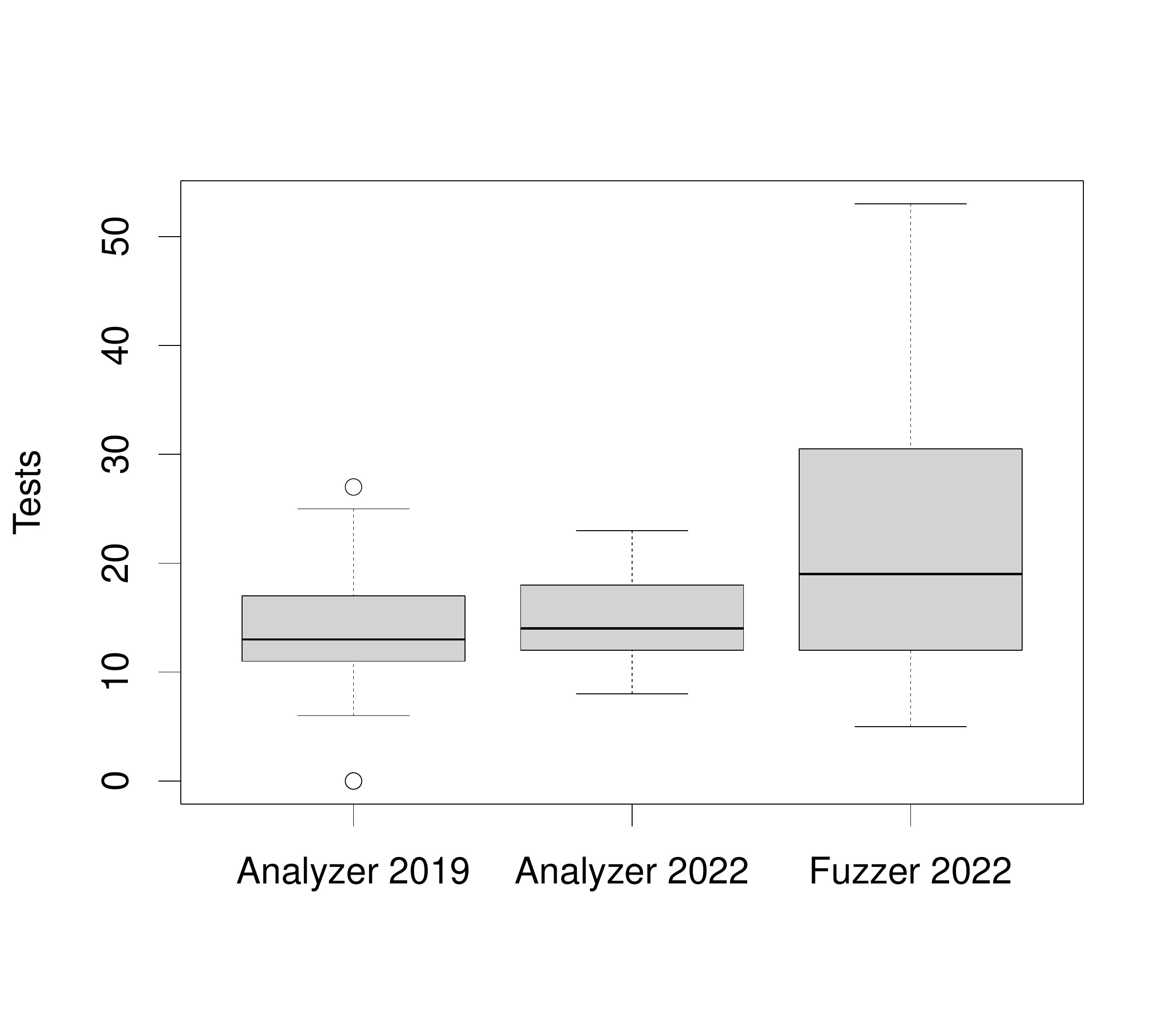}
		\vspace{-2em}
		\caption{Number of tests}
		\label{fig:tests}
	\end{subfigure}
	\hfill
	\begin{subfigure}[t]{0.49\linewidth}
		\centering
		\includegraphics[width=\textwidth]{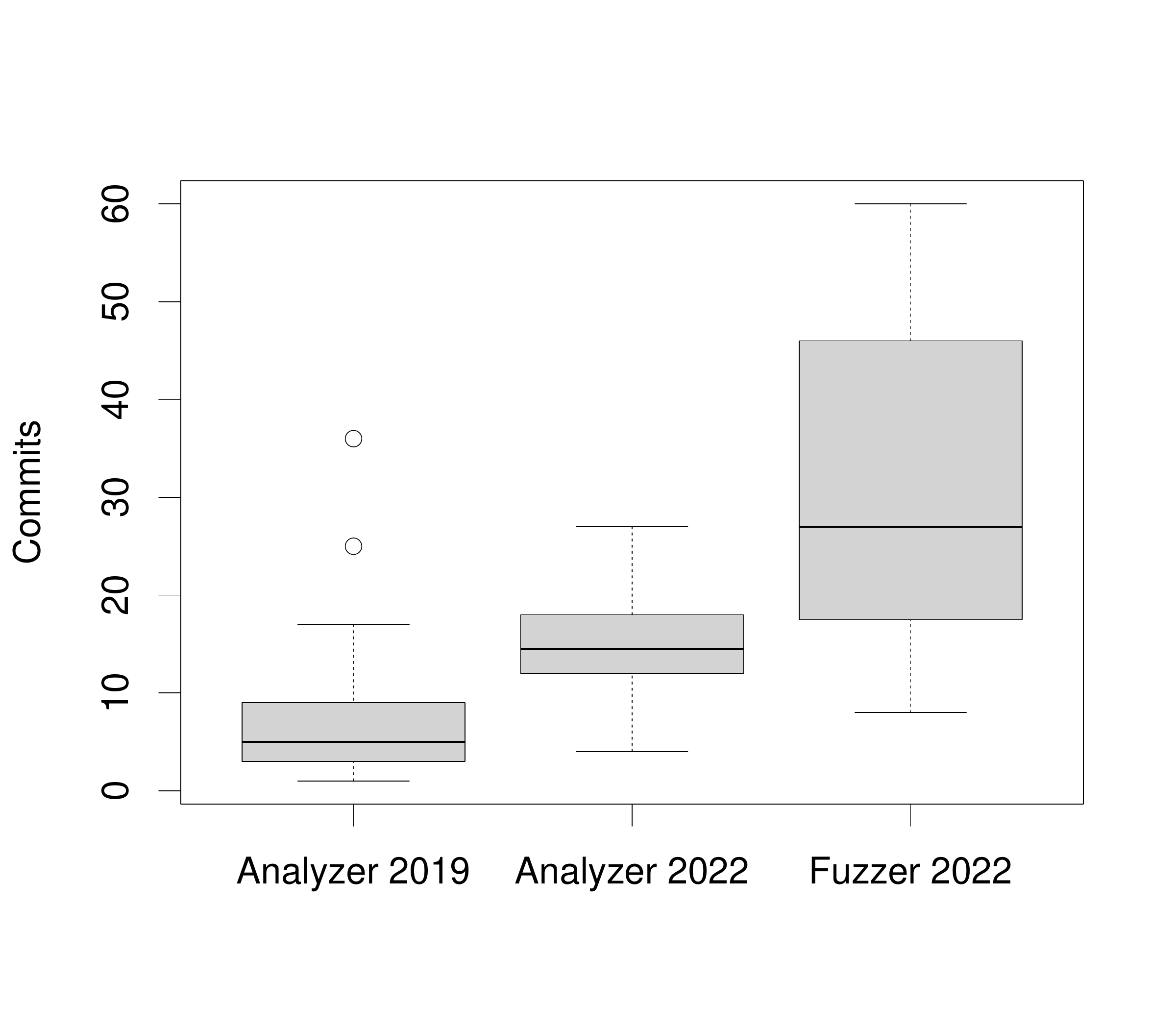}
		\vspace{-2em}
		\caption{Number of commits}
		\label{fig:commits}
	\end{subfigure}
	\hfill
	\begin{subfigure}[t]{0.49\linewidth}
		\centering
		\includegraphics[width=\textwidth]{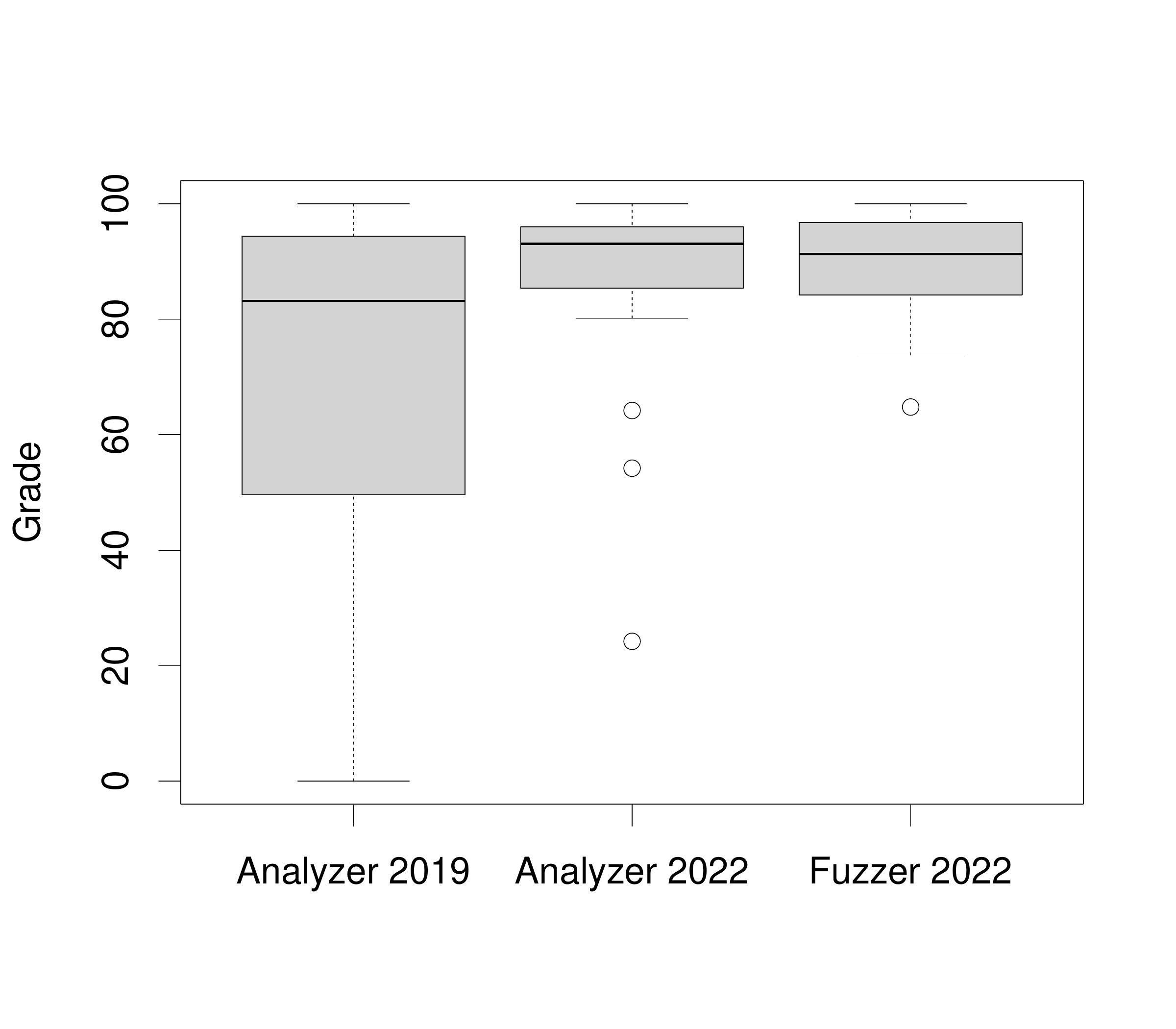}
		\vspace{-2em}
		\caption{Achieved grade}
		\label{fig:grade}
	\end{subfigure}
	\hfill
	\begin{subfigure}[t]{0.49\linewidth}
		\centering
		\includegraphics[width=\textwidth]{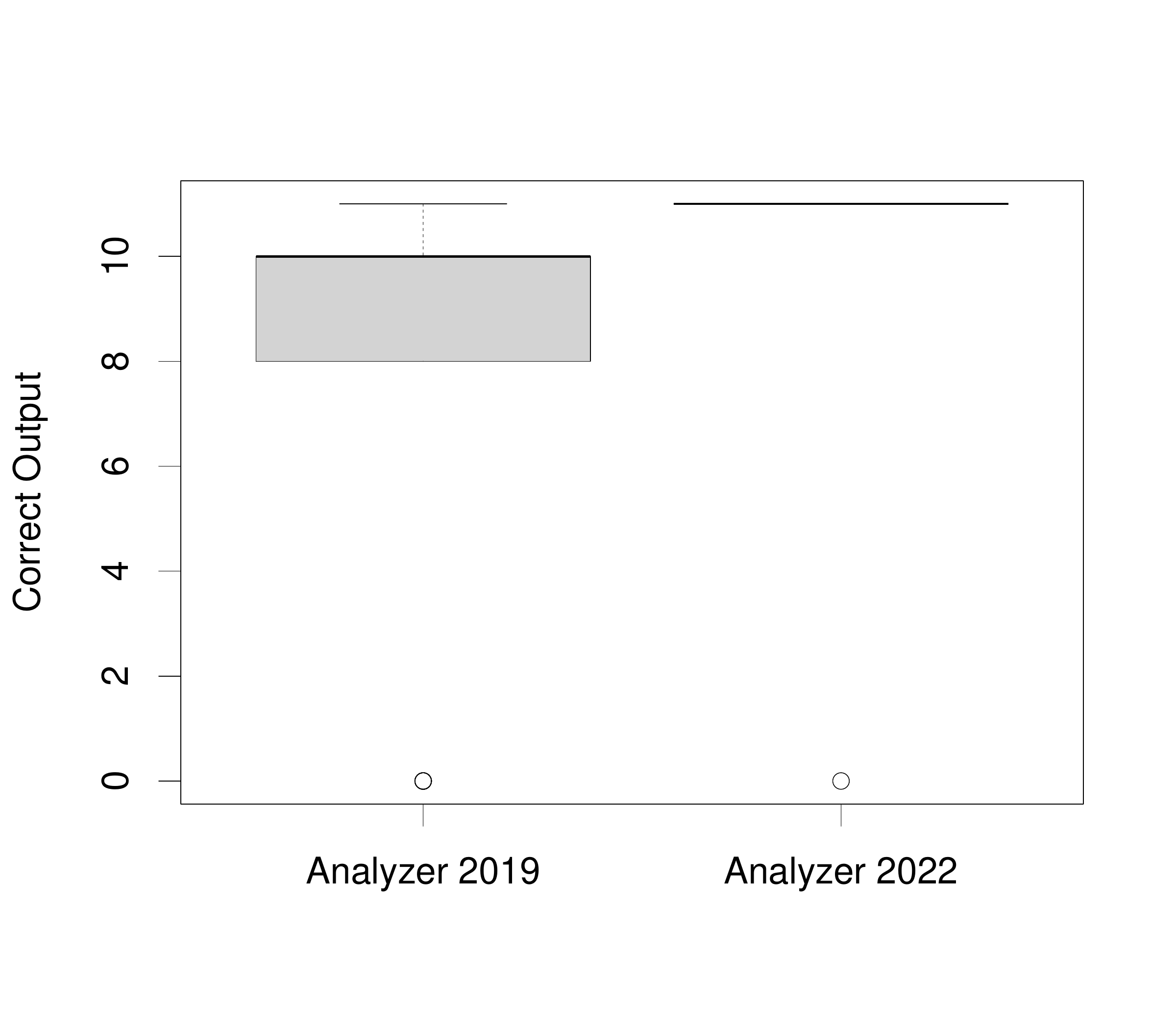}
		\vspace{-2em}
		\caption{Number of correct outputs (only available for the Analyzer)}
		\label{fig:lcaouput}
	\end{subfigure}
	
	\caption{Statistics on the testing behavior}
	\label{fig:testing}
\end{figure}

\begin{figure}
	\centering
	\begin{subfigure}[t]{0.49\linewidth}
		\centering
		\includegraphics[width=\textwidth]{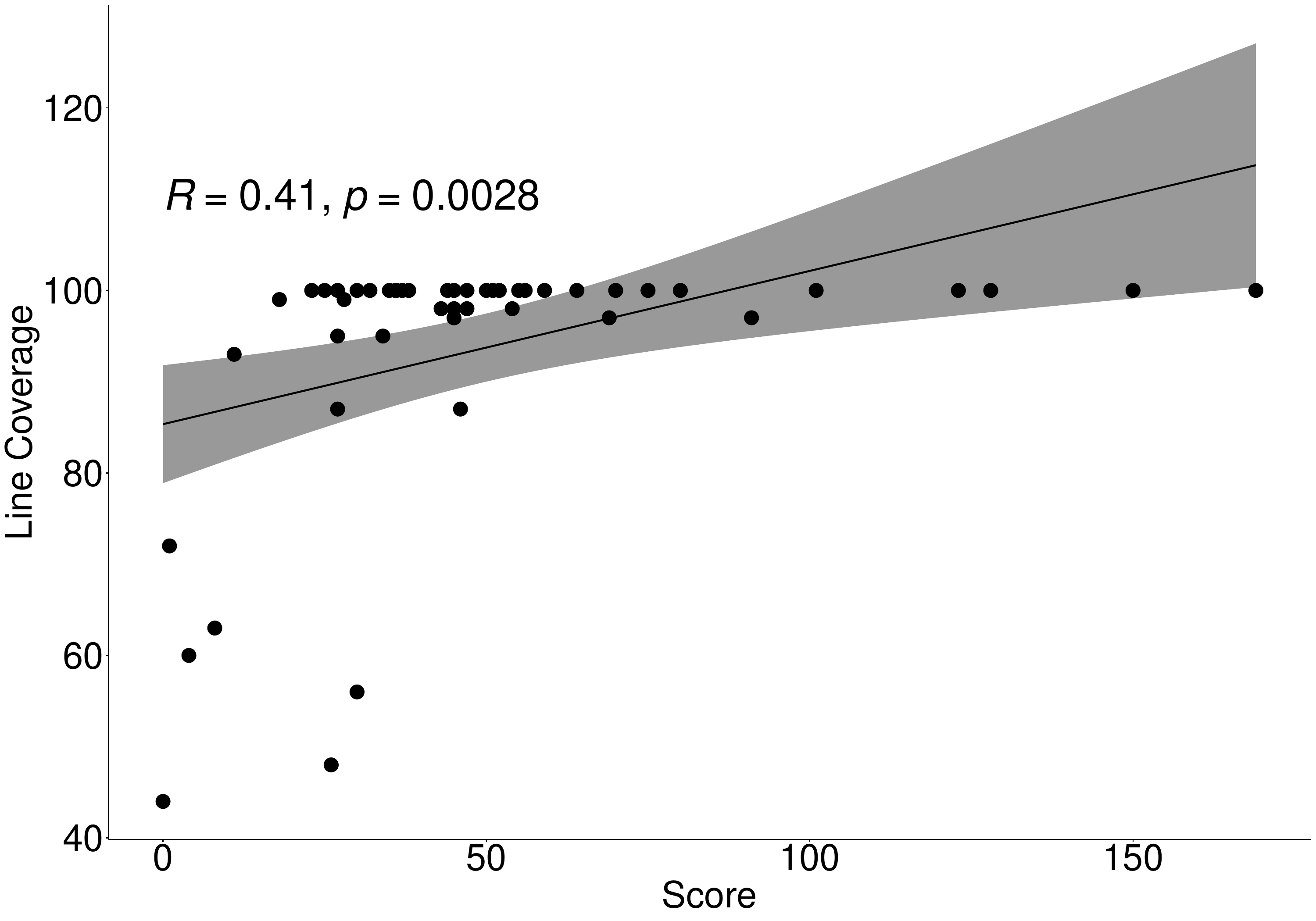}
		\vspace{-2em}
		\caption{Correlation between points and line coverage}
		\label{fig:scorecoverage}
	\end{subfigure}
	\hfill
	\begin{subfigure}[t]{0.49\linewidth}
		\centering
		\includegraphics[width=\textwidth]{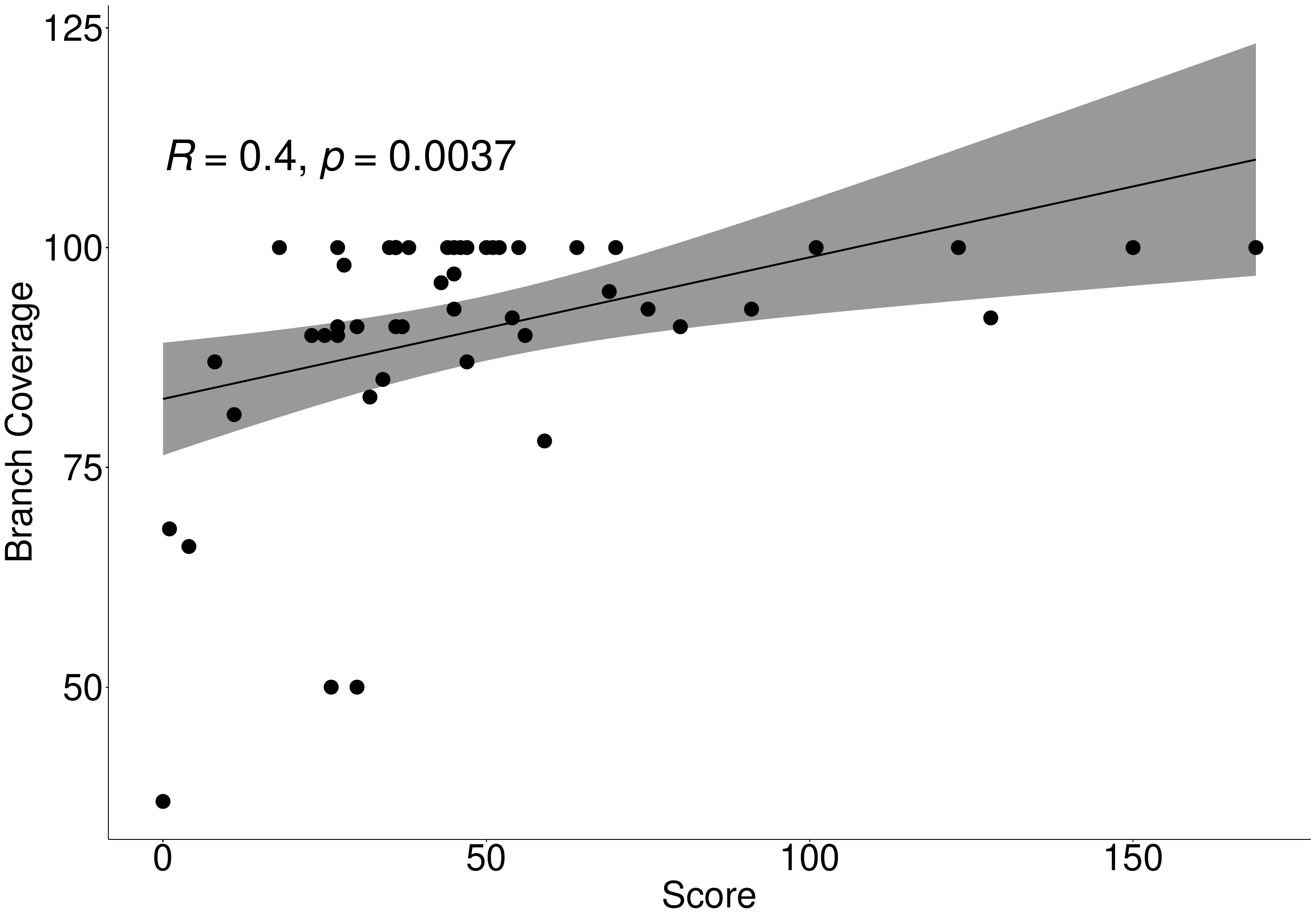}
		\vspace{-2em}
		\caption{Correlation between points and branch coverage}
		\label{fig:scorebranchcoverage}
	\end{subfigure}
	\hfill
	\begin{subfigure}[t]{0.49\linewidth}
		\centering
		\includegraphics[width=\textwidth]{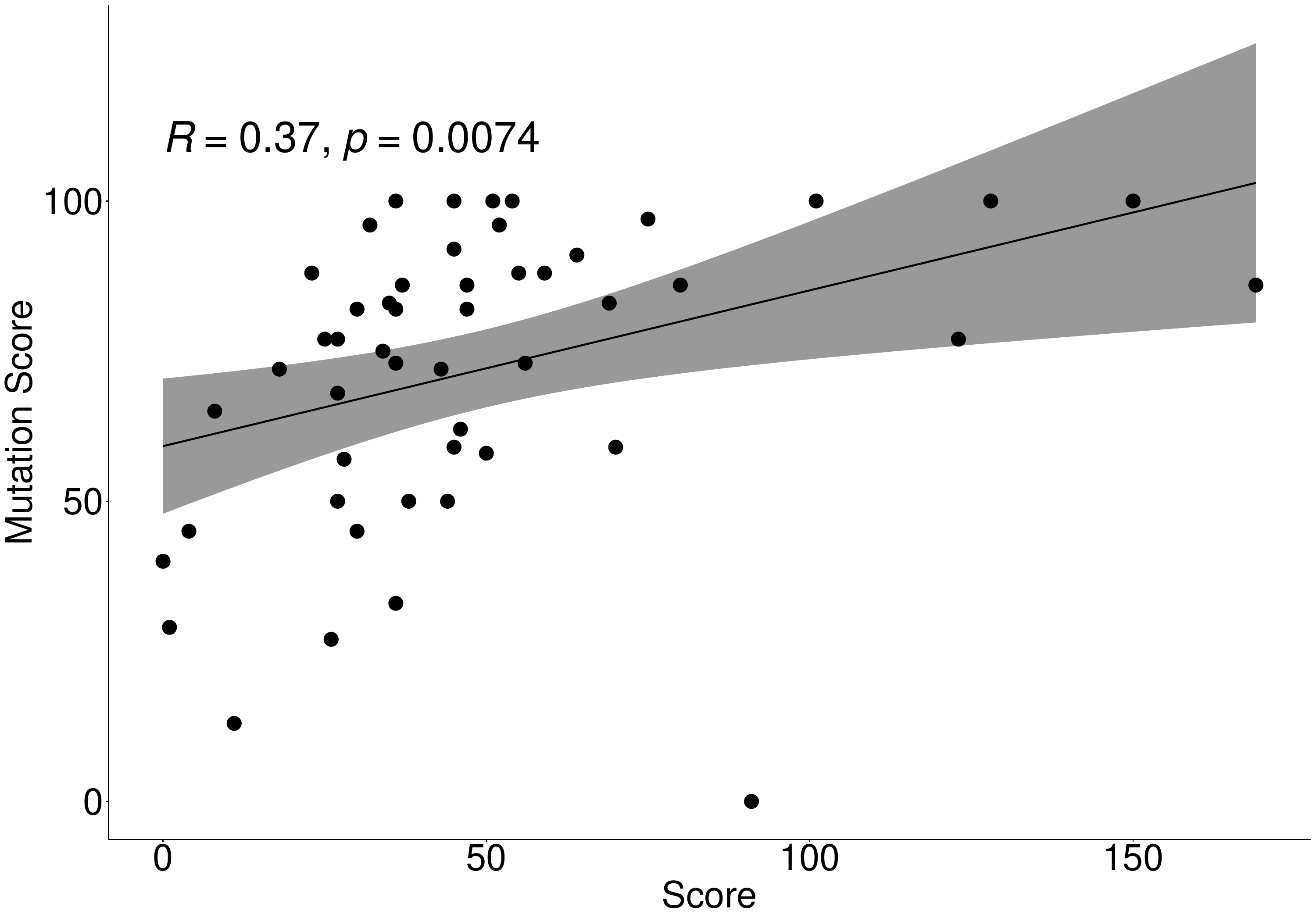}
		\vspace{-2em}
		\caption{Correlation between points and mutation score}
		\label{fig:scoremutation}
	\end{subfigure}
	\hfill
	\begin{subfigure}[t]{0.49\linewidth}
		\centering
		\includegraphics[width=\textwidth]{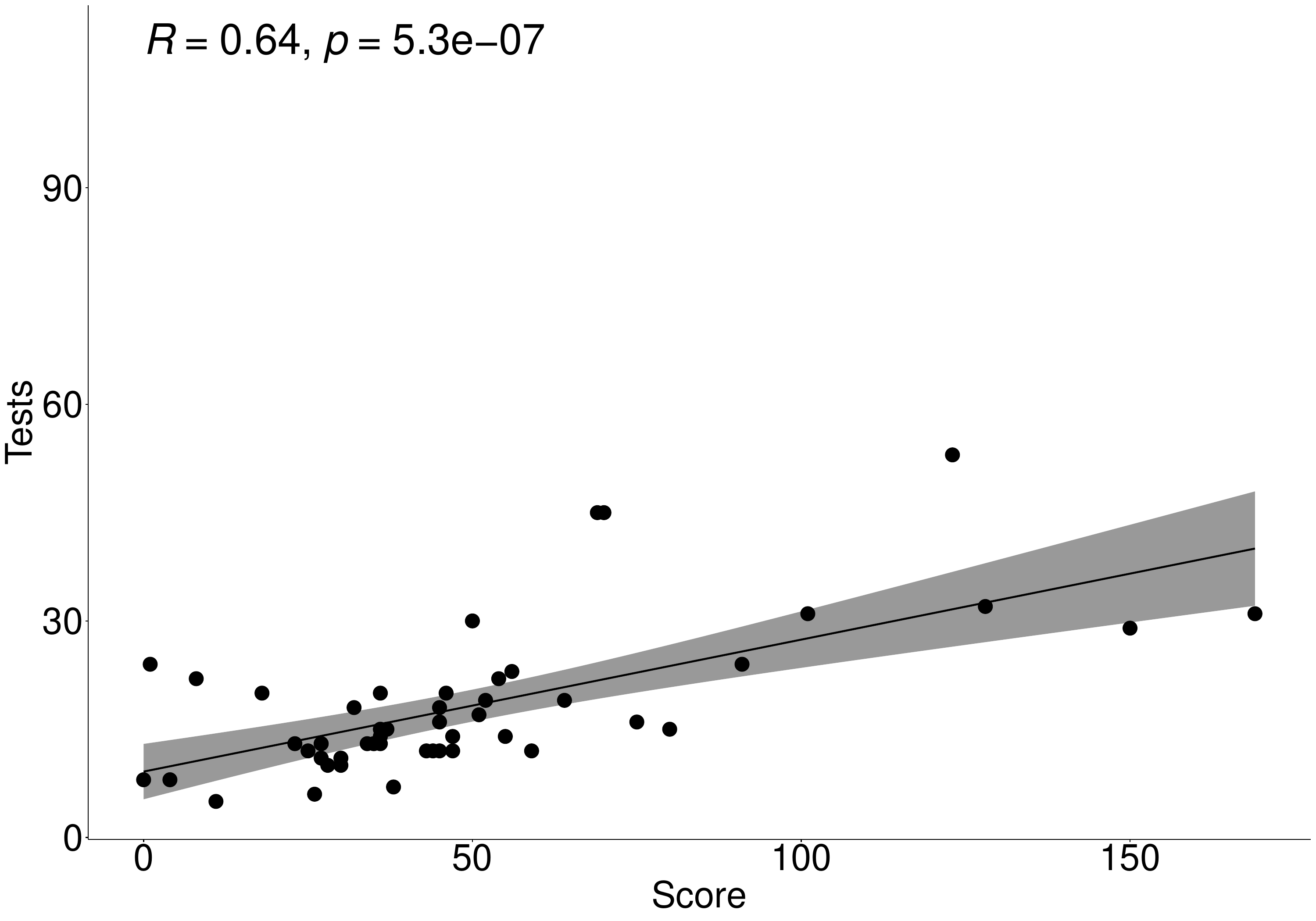}
		\vspace{-2em}
		\caption{Correlation between points and number of tests}
		\label{fig:scoretests}
	\end{subfigure}
	\hfill
	\begin{subfigure}[t]{0.49\linewidth}
		\centering
		\includegraphics[width=\textwidth]{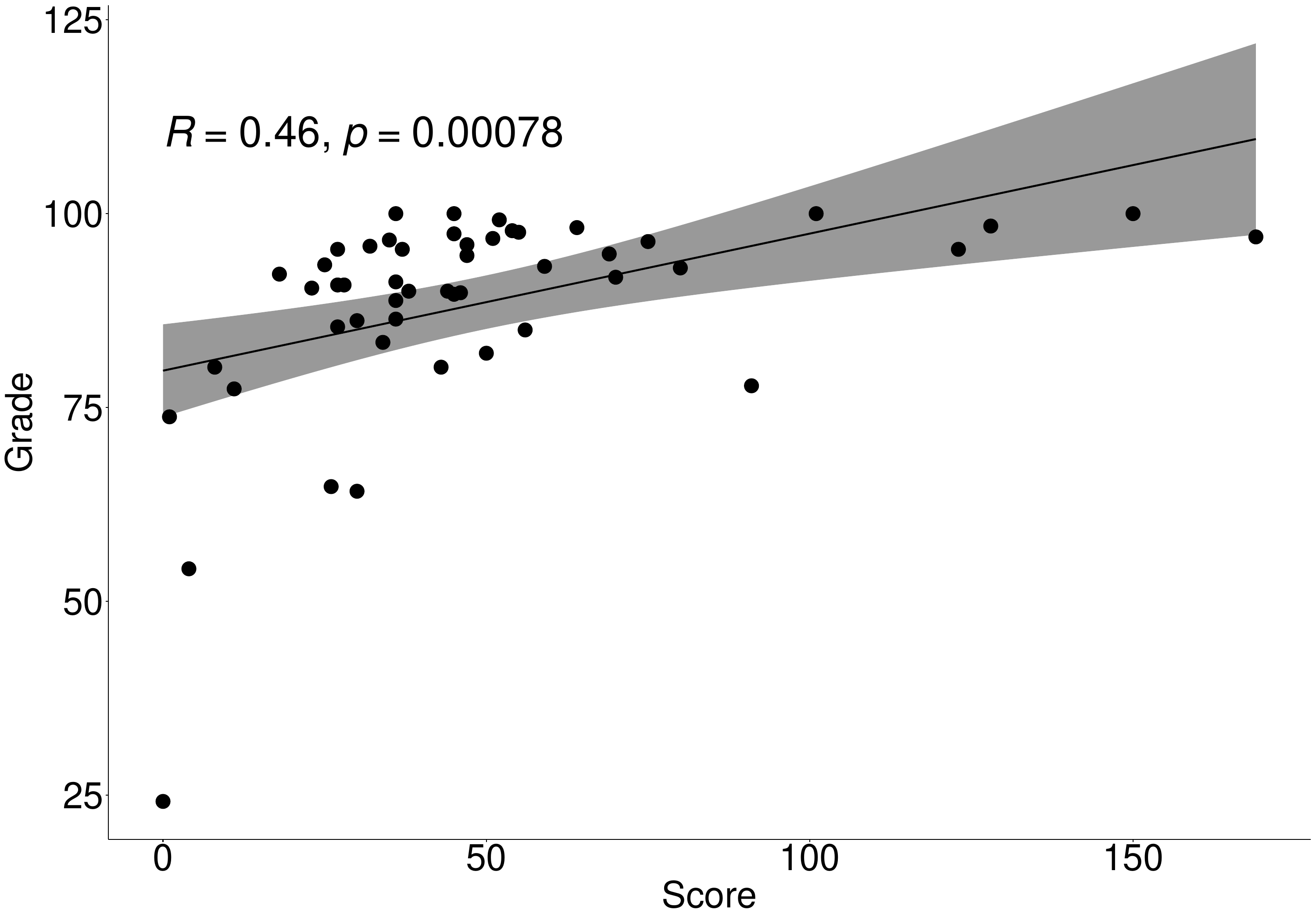}
		\vspace{-2em}
		\caption{Correlation between points and grade}
		\label{fig:scoregrade}
	\end{subfigure}
	\hfill
	\begin{subfigure}[t]{0.49\linewidth}
		\centering
		\includegraphics[width=\textwidth]{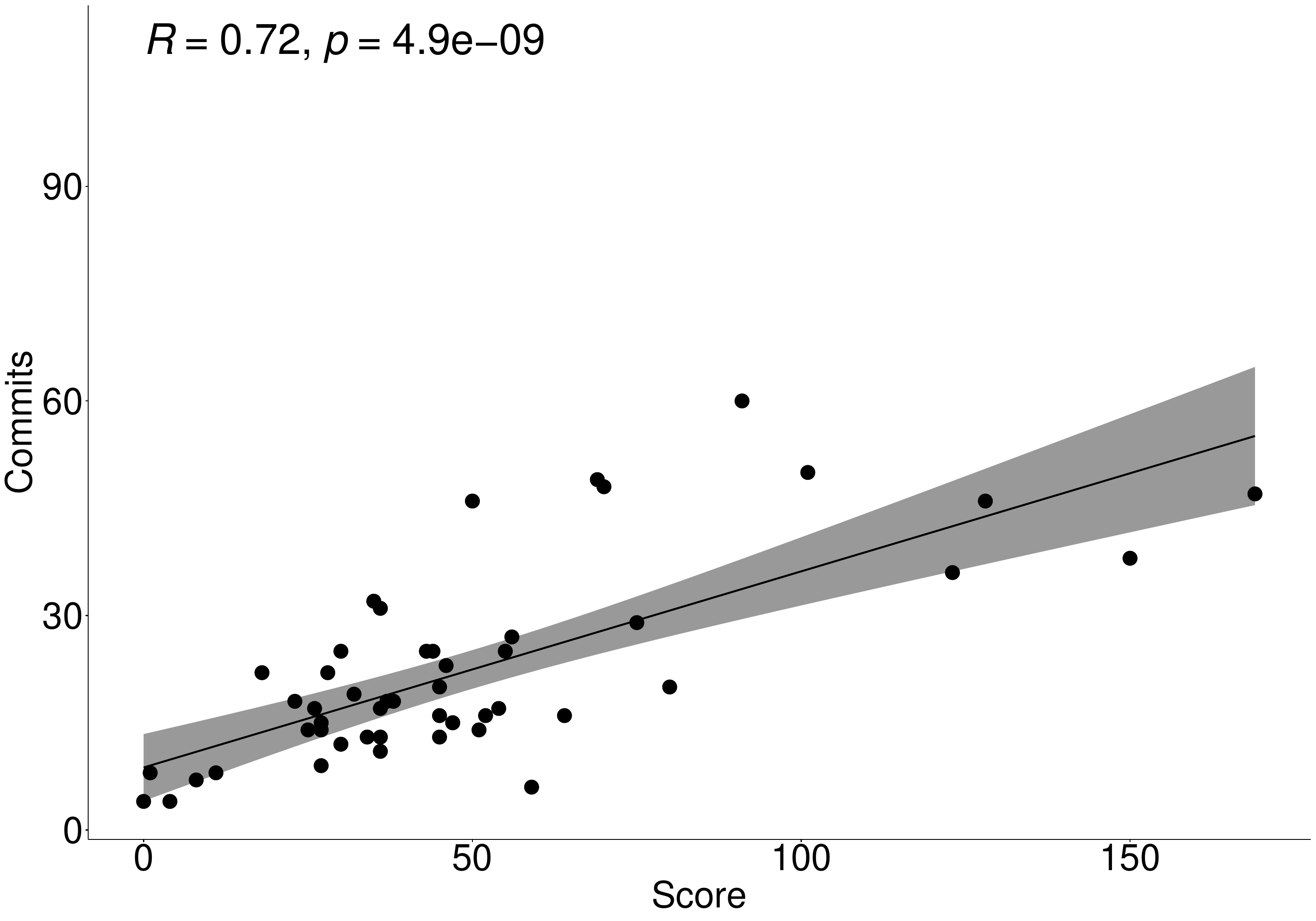}
		\vspace{-2em}
		\caption{Correlation between points and commits}
		\label{fig:scorecommit}
	\end{subfigure}
	
	\caption{Correlations between the points achieved in \toolname and different metrics of the both projects}
	\label{fig:scorecorrelations}
\end{figure}

On average, the students wrote 14.8 tests for the Analyzer project and 21.8 for the Fuzzer (\cref{fig:tests}). One of the goals of their assignments was to achieve 100\,\% line coverage (\cref{fig:coverage}) which they came close to achieving in both projects, with mean coverage percentages of 91.9\,\% and 95.5\,\% respectively.
In terms of mutation scores (\cref{fig:mutation}), the students obtained a higher score in the Analyzer (mean 78.2\,\%) compared to the Fuzzer (mean 65.2\,\%), which may suggest that implementing and testing the Fuzzer was more challenging, for example, because of the necessity to use mocking techniques and special handling of edge cases.
Despite this, the students received good grades overall for both projects (\cref{fig:grade}), with mean grades of 87.4\,\%, and 89.6\,\% respectively.
The students executed their projects in the CI more than 1300 times,
averaging almost 26 runs per student. The majority of these runs were successful (93\,\%). The high number of runs can be attributed to the number of commits of both projects, as depicted in \cref{fig:commits} (15 for Analyzer and 30 for Fuzzer).

There are significant correlations between the points accumulated by participants using \toolname and various factors, including coverage (\cref{fig:scorecoverage} and \cref{fig:scorebranchcoverage}), mutation score (\cref{fig:scoremutation}), number of tests (\cref{fig:scoretests}) and commits (\cref{fig:scorecommit}), and grade (\cref{fig:scoregrade}) for both projects combined.
These positive correlations demonstrate that engaging with \toolname and positive testing behavior are indeed related, which is important since (1) it confirms that the gamification elements are meaningful and do not distract students, and (2) it suggests that challenges and quests are effective metrics for comparing and grading students in a software testing course.
%
%

\begin{figure}[t]
	\centering
	\includegraphics[width=0.6\linewidth]{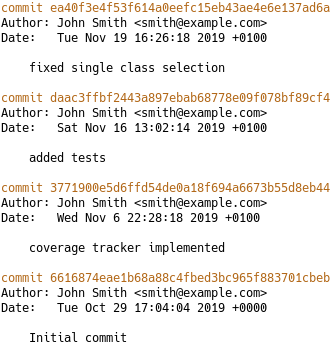}
	
	\caption{Complete \emph{git log} of a student from 2019 (without \toolname)}
	\label{fig:gitbad}
\end{figure}

\begin{figure}[t]
	\centering
	\includegraphics[width=0.6\linewidth]{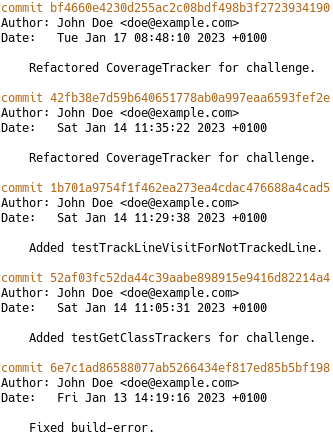}
	
	\caption{Excerpt \emph{git log} of a student from 2022 (with \toolname)}
	\label{fig:gitgood}
\end{figure}

Compared to the class of 2019 which did not use \toolname, both classes performed similarly in terms of the number of tests (\cref{fig:tests}) and line coverage (\cref{fig:coverage}). The mutation scores were higher in 2022 (\cref{fig:mutation}), although the difference is not statistically significant ($p=0.11$).
The slight increase may be because students from 2019 stopped testing their projects once they achieved 100\,\% coverage, while the students from 2022 continued to solve Mutation Challenges even after reaching high coverage.
There is a significant difference ($p<0.001$) regarding the correctness of the outputs of the Analyzer (\cref{fig:lcaouput}): In 2022, almost all students had correct outputs (mean 10.6), whereas the students from 2019 had a lower mean of 8.3. This suggests that using \toolname helped to identify bugs leading to incorrect output.
This conjecture is supported by the responses from the survey (G15), where two-thirds of the students stated that they discovered bugs while using \toolname.

The commit histories also provide evidence of bugs being found, with individual commits referring to bugs being fixed. More generally, in the 2022 course, the students committed their code in a more fine-grained manner (\cref{fig:commits}), which is evidence for their interactions with \toolname. The mean number of commits in 2022 was 15, compared to 7 in 2019; this difference is statistically significant ($p<0.001$). 
%
%
\Cref{fig:gitgood} shows an example commit history (\emph{git log}) of a student of 2022 committing tests individually, whereas \cref{fig:gitbad} exemplifies the common behavior from 2019 without \toolname when students committed their code in only a few commits.
Overall, these findings suggest that the students in 2022 were more diligent in committing their code and addressing bugs.

\summary{RQ2}{Students engaged with writing tests and achieving high coverage and mutation scores, which is reflected by metrics as well as interactions with \toolname. Compared to students not using \toolname on the same projects, we find that \toolname increased correct outputs and caused a higher frequency of commits.}

\subsection{RQ3: How did the students perceive the integration of \toolname into their projects?}

\begin{figure}
	\includegraphics[width=\linewidth]{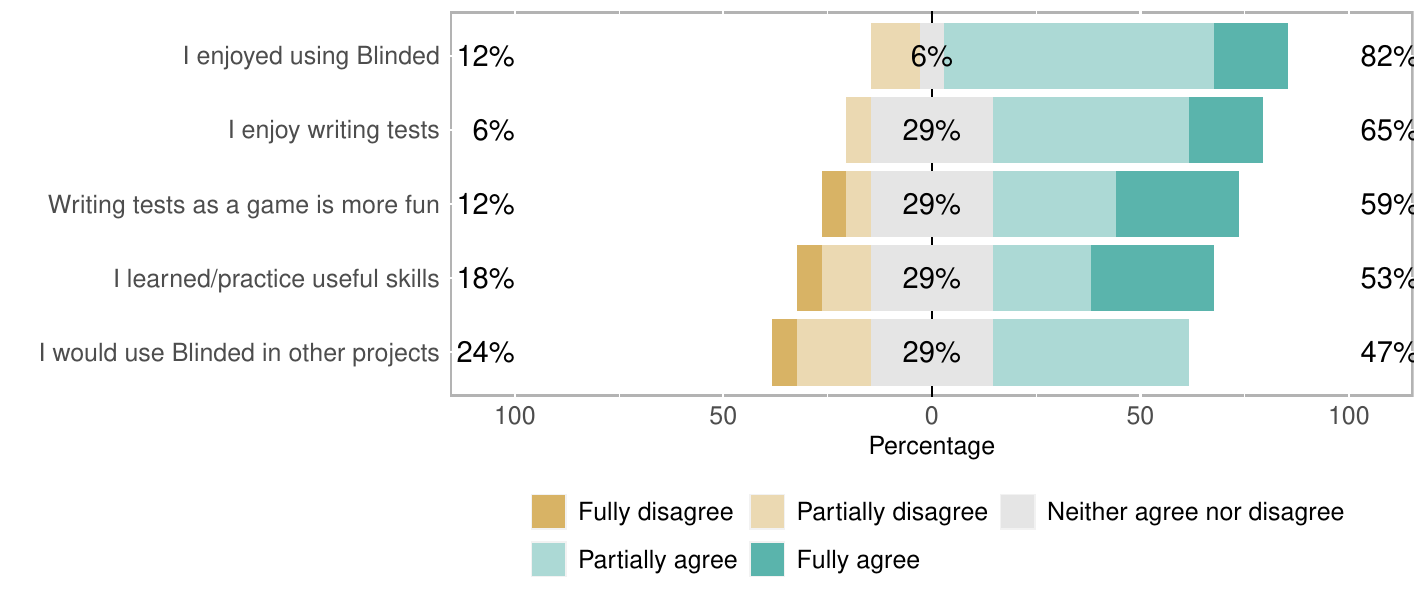}
	\caption{Responses regarding \toolname}
	\label{fig:gamekinssurvey}
\end{figure}

\begin{figure}
	\includegraphics[width=\linewidth]{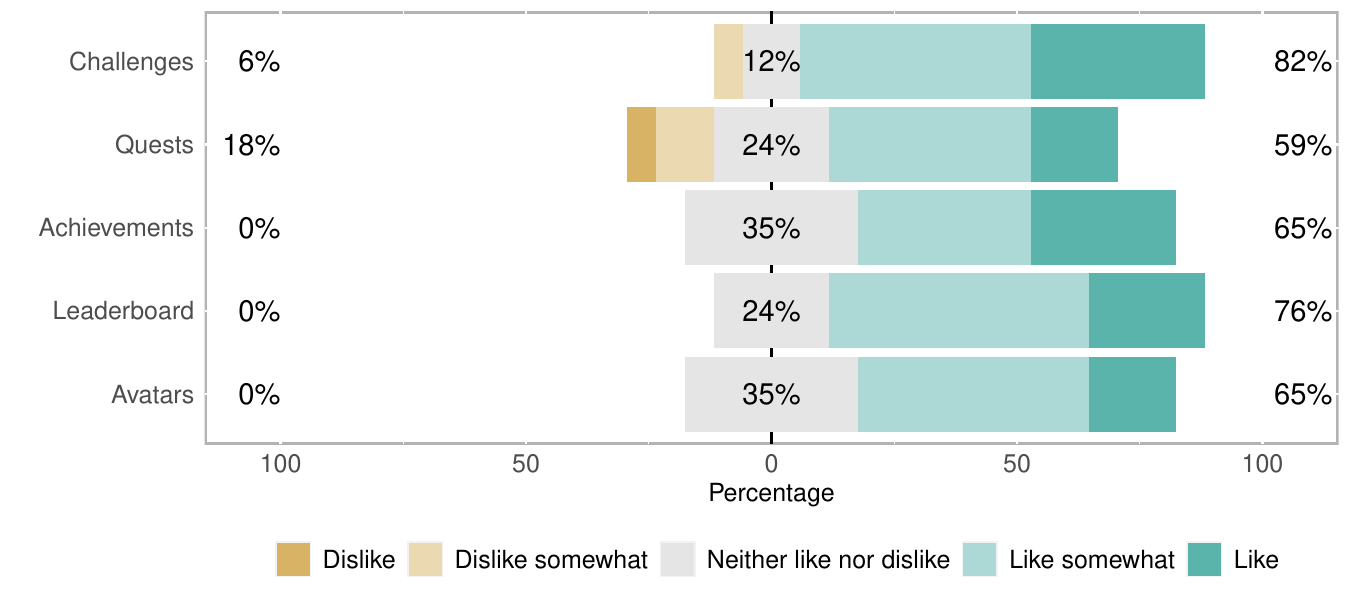}
	\caption{Responses regarding gamification elements}
	\label{fig:gamekinselements}
\end{figure}

\Cref{fig:gamekinssurvey} summarizes the survey responses: 82\,\% of the students enjoyed the inclusion of \toolname (G1) in the course, but they expressed substantially less enjoyment of writing tests (G2) in general (65\,\%). Furthermore, 59\,\% of the participants agreed that it is more enjoyable to write tests when using \toolname compared to writing tests without it (G3). Approximately half of the students reported learning and practicing valuable skills for their future careers (G4) and expressed interest in using \toolname in other projects (G5).

Considering the questions on the different gamification elements (\cref{fig:gamekinselements}), challenges were the most enjoyable aspect (G6--G10) for the majority of students (82\,\%). Quests, on the other hand, received the most negative feedback, with 18\,\% of students expressing dissatisfaction. 
The sequential nature of solving the steps of quests appears to be difficult, especially since the application being tested was small and had only few classes. This frustrated some students who found it challenging to progress without solving multiple steps at once. However, students appreciated the competitive aspect of the leaderboard and enjoyed the challenges and achievements.

Approximately 24\,\% of the students reported feeling pressured to solve challenges (G12), with all of them expressing the desire to achieve a good grade in the course. While this likely holds for any required aspect of coursework, it raises the question of how well practitioners would engage with \toolname.
While a majority of the students (59\,\%) stated that they wrote more tests using \toolname (G13), an equal number of participants (59\,\%) mentioned that they believed they wrote unnecessary tests just to solve challenges (G14). Unfortunately, no reasons were provided by these students. One possible explanation could be that they were required to achieve 100\,\% coverage for their project, and any tests that went beyond that (such as those targeting mutants) were perceived as unnecessary. On the other hand, it may simply be the case that the 100\,\% coverage requirement implied coverage on trivial code (e.g., getters or setters). However, neither of these reasons would be specific to \toolname, but would equally apply to the underlying test metrics.

The students encountered several problems (F2--F4):
Issues arose when they had to modify code such that this caused \toolname to fail to locate relevant lines after the changes, thus automatically rejecting challenges. Sometimes Mutation Challenges were solvable on the students' computers but not in \toolname, which was caused by changes in line numbers, modified code, or PIT not recognizing the mutant as killed. Smell Challenges also caused some confusion whenever it was not apparent without further explanation why a suggested change would improve the code in the given context. The students also faced problems with Quests, which sometimes were impossible to complete fully due to code changes. These issues will be addressed in future versions of \toolname,  for example by improving the tracking of code changes, rethinking the concept of Quests, and reducing the number of smells reported. 

Overall, however, the students expressed satisfaction with the integration of \toolname into the course (F1). They appreciated the small tasks assigned to them instead of being overwhelmed by a large number of missed scenarios. They particularly enjoyed the competitive aspect and the game-like interface, which provided hints and ideas for testing and improving tests and code.

\summary{RQ3}{Students appreciated the integration of \toolname and the implementation of tests in their coursework. They particularly enjoyed the challenges, but were less fond of quests. Some students felt they wrote more tests than necessary.} 


%% file: sections/relatedwork.tex
Gamification has been explored for incentivizing practitioners to write tests and adopt good testing practices~\cite{Fulcini2023}. For this, gamification has been integrated into integrated development environments~\cite{blinded}, browsers~\cite{Fulcini2022}, and continuous integration~\cite{Ayoup_2022}, with results suggesting that gamification can improve test automation. There have also been attempts to gamify aspects of testing such as test-to-code traceability~\cite{DBLP:conf/sera/Parizi16} or acceptance testing~\cite{DBLP:conf/icse/ScherrEH18}. While we also address continuous integration, our gamification approach aims to influence testing behavior directly, rather than the adoption of tools, and we use more different gamification elements.

Gamification has also been applied to software testing education, incorporating different gamification elements and concepts. Unplugged approaches use tools like Lego or dice to teach testing concepts~\cite{Lorincz2021}. Others utilize virtual races~\cite{Blanco2023}, quizzes~\cite{DBLP:conf/sbqs/JesusPFS19}, business trainings~\cite{DBLP:conf/icsob/Yordanova19}, or board games~\cite{Moreira2022}. Gamification has also been applied to tutorial systems for teaching software testing~\cite{elbaum2007bug, fu2016gamification} and to mutation testing~\cite{fraser2019gamifying}. Additionally, serious games have been developed to teach software testing~\cite{DBLP:conf/icst/StraubingerCF23, bell2011secret}. Most of these studies report that the use of gamification had a positive impact on testing, and our approach adds further evidence that this is the case. In contrast to prior work, our approach differs as gamification is integrated into the development process via CI, and it uses different gamification elements.
%
However, gamification elements do not universally resolve the lack of motivation; a recent systematic mapping study identified negative effects such as lack of motivation and effectiveness in some cases~\cite{Almeida2023}. In our study, such effects mainly occurred when there were technical issues with \toolname.


%% file: sections/conclusion.tex
Overall, the integration of the \toolname plugin for Jenkins into the software testing course proved to be successful in engaging students in testing activities that go beyond formal requirements for the course using gamification elements such as challenges, quests, achievements, leaderboards, and avatars. This resulted in a significant improvement in the correctness of the programs they wrote as part of their coursework, and generally very positive feedback.


Our study revealed several opportunities to further improve \toolname and its integration into a testing course. Specifically, the quests, challenges, and achievements should be reworked and refined to address issues such as when they are unsolvable with a better tracking of modified code. Additionally, an interesting suggestion made by students is to incorporate an IDE plugin to display challenges directly within the development environment.
Ongoing work on \toolname also includes the adoption of new programming languages and testing frameworks. While \toolname has been effective in a university setting, evaluating its potential in an industry setting would also be valuable. Additionally, an evaluation using more complex student projects could provide valuable insights into team behavior and longer-term effects of \toolname.
